\documentclass[12pt,preprint]{aastex}
\received{2003 August 6} \slugcomment{Accepted for ApJS August
2004, v153 issue.}
\begin{document}
\title{W49A North - Global or Local or No Collapse?}
\author{John A.  Williams\altaffilmark{1}}
\affil{Physics Department, Albion College, Albion, MI 49224}
\altaffiltext{1}{Department of Astronomy, University of Illinois, 103 Astronomy Building,
1002 West Green Street, Urbana, IL 61801}
\email{jwilliams@albion.edu}
\author{H{\'e}l{\`e}ne R.  Dickel}
\affil{Department of Astronomy, University of Illinois, 103 Astronomy Building,
1002 West Green Street, Urbana, IL 61801}
\email{lanie@astro.uiuc.edu}
\and
\author{Lawrence H.  Auer}
\affil{Earth and Environmental Sciences 5, Los Alamos National Laboratory,
MS-F665, Los Alamos, NM 87545}
\email{lhainnm@mindspring.com}

\begin{abstract}
We attempt to fit observations with 5\arcsec~resolution of the
J=2$-$1 transition of CS in the directions of H~{\sc ii} regions
A, B, and G of W49A North  as well as observations with 20\arcsec~
resolution of the J=2$-$1, 3$-$2, 5$-$4, and 7$-$6 transitions in
the directions of H~{\sc ii} regions A and G by using radiative
transfer calculations. These calculations predict the intensity
profiles resulting from several spherical clouds along the line of
sight. We consider three models: global collapse of a very large
(5 pc radius) cloud, localized collapse from smaller (1 pc) clouds
around individual H~{\sc ii} regions, and multiple, static clouds.
For all three models we can find combinations of parameters that
reproduce the CS profiles reasonably well provided that the
component clouds have a core-envelope structure with a temperature
gradient. Cores with high temperature and high molecular hydrogen
density  are needed to match the higher transitions (e.g. J=7$-$6)
observed towards A and G. The lower temperature, low density gas
needed to create the inverse P-Cygni profile seen in the CS
J=2$-$1 line (with 5\arcsec~ beam) towards H~{\sc ii} region G
arises from different components in the 3 models.  The infalling
envelope of cloud G plus cloud B creates the absorption in global
collapse, cloud B is responsible in local collapse, and a separate
cloud, G', is needed in the case of many static clouds. The exact
nature of the velocity field in the envelopes for the case of
local collapse is not important as long as it is in the range of 1
to 5 km~s$^{-1}$ for a turbulent velocity of about 6 km~s$^{-1}$.
High resolution observations of the J=1$-$0 and 5$-$4 transitions
of CS and C$^{34}$S may distinguish between these three models.
Modeling existing observations of HCO$^+$ and C$^{18}$O does not
allow one to distinguish between the three models but does
indicate the existence of a bipolar outflow.
\end{abstract}


\section{INTRODUCTION}

\objectname{W49A}
North is a giant H~{\sc ii}-molecular cloud complex at a distance of 11.4 kpc
\citep{gwi92}\footnote{Gwinn et al. refer to W49A North as W49N.}.  One of the
earliest observations and models of W49A was made by \citep{muf77}.  Although they
do not distinguish between W49A North and W49A South, their figures 3 and 4 indicate
the emission they observed came mainly from W49A North.  With resolutions of
1\arcmin~to 2\arcmin, they detected an H~{\sc ii} region in the H76 alpha line at
about 8 km~s$^{-1}$ and molecular clouds in the 2 cm line of formaldehyde at 4
km~s$^{-1}$ and 12 km~s$^{-1}$ with respect to the local standard of rest. Because
the formaldehyde observations indicated that the H~{\sc ii} region is seen through
the higher-velocity cloud, their picture is of an H~{\sc ii} region between two
molecular clouds with the one on the near side of the H~{\sc ii} region moving away
from us at 12 km~s$^{-1}$ and the one on the far side moving away at 4 km~s$^{-1}$.

\citet{wel87} detected what appeared to be a ring of H~{\sc ii}
regions\footnote{High resolution images and nomenclature of the
H~{\sc ii} regions A through M in this ``ring'' of H~{\sc ii}
regions of W49A North are given by \citet{dre84} and \citet{DG90}
and by \citet{dwgwm00} for their subcomponents.} seen obliquely in
the radio continuum emission at 6 cm.  Their observations with the
Hat Creek mm-array of the J=1$-$0 transition of HCO$^+$ with a 7
arcsecond beam centered on the 3-mm continuum maximum showed
absorption of the continuum emission on the high velocity side of
the profile which they interpreted as due to collapse motion of
gas toward a central mass. This was their primary argument against
the earlier two-cloud interpretation of double emission peaks seen
in the J=1$-$0 lines of CO (e.g. \citet{sco73}, \citet{muf77}, and
\citet{miy86}).

On the basis of observations of several transitions of CS with 12 to 20 arcsecond
beams, \citet{ser93} argue that the double peaks can not be due to self-absorption
because their observations of C$^{34}$S show the peaks in lines that should not be
saturated because the terrestrial abundance of $^{34}$S is 23 times smaller than the
abundance of the ordinary isotope, $^{32}$S.  They also argue that the emission at
low velocity (near 4 km~s$^{-1}$) and absorption at high velocity (near 12
km~s$^{-1}$) seen towards the H~{\sc ii} region G are due to separate molecular
clouds, not the collapse of one cloud as Welch et al. proposed, because the CS
emission from the 12 km~s$^{-1}$ gas is seen over a large part of the field and
stays at 12 km~s$^{-1}$ rather than dropping to, e.g., 8 km~s$^{-1}$ at the edges.
Serabyn et al. suggest  that the two clouds are colliding, and the collision has
triggered enhanced O-star formation.

\citet{DA94} using a multi-level, non-LTE radiative transfer code
found that a large-scale free-fall collapse reproduced the HCO$^+$
J=1$-$0 inverse P-Cygni profile observed with 7\arcsec~resolution
toward H~{\sc ii} region G and the J=3$-$2 profiles observed with
24\arcsec~resolution at locations both towards and away from G.

\citet{ket91}, using a hydrodynamic simulation of gravitational
fragmentation, concluded that Welch et al.'s ``ring of H~{\sc ii}
regions'' may have formed in the gravitational fragmentation of a
flattened, rotating molecular cloud.  In contrast to a global
collapse, Keto et al. say that the double-peaked emission lines
and absorption lines and even the inverse P-Cygni profiles are
indicative of localized accretion flows in individual star-forming
fragments.

We undertook the present study to see if we could find a way to distinguish between
these competing models: multiple static clouds, global collapse, or local collapse, using
the radiative transfer code of Dickel \& Auer and observations with the
Berkeley-Illinois-Maryland-Association (BIMA) array of the CS J=2$-$1 transition
with a 5 arcsecond beam (Dickel et al. 1999, paper 1) together with the
lower-resolution observations of Serabyn et al.

\section{TOOLS}

We used two Fortran programs, rt and mc, to create model molecular
clouds and to compare intensity profiles one would observe from
them with observations. The multi-level, non-LTE radiative
transfer code, rt, which is based on an accelerated lambda
iteration method is described in the appendix of \citet{DA94}.  It
assumes spherical symmetry in the cloud, but the variation of
density, molecular abundance, velocity, and other parameters with
radial distance is arbitrary.  The program automatically handles
regions where there is a transition to or from population
inversions.  The program creates files of emergent intensity and
optical depth as functions of impact parameter, velocity, and
transition.

These files are used by the display program, mc.  This program
(described more fully in the Appendix) assumes that the radiation
from one cloud does not affect the populations of the energy
levels in another cloud, so each cloud just attenuates the
radiation from clouds behind it and adds its own contribution to
the intensity.  The validity of this assumption is discussed in
the Appendix.  For each cloud the user specifies the velocity with
respect to the local standard of rest, the offset on the sky from
some origin, and the relative position along the line of sight.
The number of clouds that can be included is limited only by the
size of the memory of the computer and the patience of the user.
The outputs include a graphical display for the Tektronics window
under X Windows, a file with a postscript version of the display,
a log file with information about the display and the rms
difference between the model profile and the observed profile, and
a file with the computed profiles.

\section{OBSERVATIONS}

We used observations of several transitions of the CS molecule in
the directions of H~{\sc ii} regions A and G and in the direction
of the central cloud of \citet{ser93}.  The data consist of
observations of the J=2$-$1 transition of CS convolved to
5\arcsec~resolution in the directions of H~{\sc ii} regions A, B,
and G with the Berkeley-Illinois-Maryland-Association (BIMA)
array.  Those in the directions of A and G were also convolved to
20\arcsec~to compare with the Serabyn et al. observations.  The
intensities as a function of velocity for the CS emission from the
J=3$-$2, 5$-$4, and 7$-$6 transitions of CS observed with
20\arcsec~resolution in the directions of Serabyn et al.'s central
clump and southwestern clump were read from Figure 2 of their
paper. The BIMA observations are described more fully in paper 1.

There are a number of features that models of the complex should
reproduce. Serabyn et al. saw an extended region of CS emission
elongated from southwest to northeast with a velocity gradient
which they interpreted as three partially overlapping clumps. When
convolved to the same resolution (12\arcsec$-$20\arcsec), the BIMA
observations of the J=2$-$1 CS line show the same features.  At
higher resolution (5\arcsec) we see inverse P-Cygni profiles at G
and a ``C'' shape in P$-$V plots through G.  These features are
not seen at lower resolution.  We consider only the southwestern
and central clumps in Serabyn et al.'s model.  There are a number
of H~{\sc ii} regions in the complex.  We consider H~{\sc ii}
regions A, B, and G which are near the southwest and center of the
complex, and attempt to reproduce the observed CS profiles and
line strengths in these directions.  In the high resolution CS
J=2$-$1 spectra where the continuum emission has not been removed
(Figure 10a of paper 1), the profiles in the directions of A, B,
and G have similar intensities between 10 and 20 km~s$^{- 1}$.

We converted the observed brightness temperatures T$_B$(K) to intensities
I (ergs~s$^{-1}$~cm$^{-1}$ Hz$^{-1}$ ster$^{-1}$) when displaying the observed
and theoretical line profiles. The relationship between T$_B$(K)
and I(cgs) is given by:
\begin{equation}
   T_B(K) = {c^2 \over {(2 k \nu^2)}} ~10^{14}~ I(cgs) = C(\nu)~ 10^{14}~ I(cgs)
\end{equation}
where c is the velocity of light, k is Planck's constant, and
$\nu$ is the frequency in Hz.
The values of C($\nu$) are given in Table 1 for the observed transitions of CS
and C$^{34}$S.

\section{PROCEDURE}

\subsection{Overall Strategy}

The overall strategy was to start with each of the three models mentioned above,
multiple static clouds, global collapse, and localized collapse, using the parameters
suggested by their proponents.  The sizes of the H~{\sc ii} regions came from
\citet{dre84}.  The electron densities in the H~{\sc ii} regions were adjusted to
make the continuum levels predicted by the models fit the observed levels.  The
cloud densities, velocities, turbulent velocities, molecular abundances, and sizes
were then adjusted to fit all the observed line profiles as well as possible.  The
fit was judged both by eye and by the rms differences between the model profiles and
the observed profiles.  Then additional clouds were added as seemed indicated by
the observations.  Finally we relaxed the assumption of uniform conditions
within the clouds.

\subsection{Techniques and Insights}

\subsubsection{Effects of Changes in Basic Parameters}

This section will summarize the effects on a spectral profile of
changes in the basic parameters of a model cloud with an H~{\sc
ii} region at its center.

Section 5.3.2 of \citet{DA94} describes using the observed
continuum level to fix the electron density in the central H~{\sc
ii} region of the cloud.  We followed the same procedure.

For an optically thin gas which is collisionally
excited, the emission should scale with the square of the volume density (particles
~cm$^{-3}$).
Figure 4 and sections 6.1 and 6.2 of \citet{DA94} illustrate the effects of varying
the volume density of molecular hydrogen and the relative abundance of HCO$^+$ for
the J=0$-$1 and 3$-$2 transitions.  The figure shows that the strength of the line
increases with increasing molecular abundance for both transitions.  It also shows
that the higher transition is much more sensitive to the molecular hydrogen density
than is the lower transition.

As is well known, motion of the molecules due to thermal motion,
turbulence, and gradients in outflows or inflows broaden the
lines.  We combine thermal and non-thermal turbulent motion into a
microturbulent velocity.  The effects of gradients in inflows and
outflows is discussed in section 5.3.3 of \citet{DA94} in the case
where the line of sight goes through the center of a cloud that
shows a P-Cygni or inverse P-Cygni profile in one of the
transitions.  Their Figure 3 relates conditions in the clouds to
intensities in the line profiles for velocity fields representing
free-fall and homologous collapse.

Our tests show that the behavior of lines from the CS molecule is
similar to the behavior of HCO$^+$ lines. Most of the CS
transitions in our clouds are optically thin. However, the gas
becomes optically thick in the J=5$-$4 and J=7$-$6 transitions for
cloud A and in the gas causing the absorption in the J=2$-$1 line
-- which is cloud G' in the multi-cloud model, cloud B in the local collapse
model, and cloud B and the envelope of cloud G in the global collapse model.

The brightness temperature of the emission is the product of the
excitation temperature and $1 - e^{-\tau}$ where $\tau$ is the
optical depth of the transition.  As the optical depth increases,
the brightness temperature approaches the excitation temperature.
The excitation temperature of a transition depends on the kinetic
temperature, the density of colliders (hydrogen molecules) and the
energy, $E_u$, of the upper level.  The ``equivalent temperature''
in kelvins is $E_u$/k, where k is the Boltzmann constant.  When
the kinetic temperature is below the equivalent temperature of a
level, then only a small fraction of the collisions have enough
energy to populate the level, so very high densities are needed.
The J=2 level of CS is easy to excite because its equivalent
temperature is only 7K whereas the J=7 level with an equivalent
temperature of 66K is much harder to excite. The ``critical
density'' is the density for which the levels become thermalized,
i.e. the excitation temperature equals the kinetic temperature
\citep{i87}. For kinetic temperatures between 50K and 100K, this
occurs at H$_2$ densities, n(H$_2$), of $4.2\times 10^5$~cm$^{-3}$
for the J=2$-$1 transition and $2.0\times 10^7$ ~cm$^{-3}$ for the
J=7$-$6 transition. If the levels are subthermal and the
overpopulation of the upper level is small, then any variation in
the background radiation field is amplified by the same factor and
is relatively small, i.e. the emergent intensity is linearly
proportional to the background kinetic temperature. If the
populations become strongly inverted, then there is exponential
amplification and comparatively small changes in the negative
optical depth will result in exponentially larger variation in the
emergent intensity.  In the outer parts of most of our model
clouds, the CS is subthermally excited, but in the central parts,
the lower CS levels are thermalized and in some cases become
suprathermal. For some models, there is a small range of radii
where the populations of the lowest levels are inverted and weak
masering occurs in the CS J=1-0 transition.

The emission can be further enhanced when the opacity is high
enough to trap a significant fraction of the radiation.  In this
case photons are available to excite
the molecules causing the excitation temperatures to be higher
than with collisions alone. We found evidence of this radiative
trapping when we put a lower density envelope around a cloud
(section 5.1.2) and when we replaced a cloud with a smaller,
denser core surrounded by a larger, infalling envelope (section
5.3); the results were a somewhat flatter and broader distribution
of the maximum optical depths of transitions as a function of the
value of J of the upper level and excitation temperatures which
did not decline as steeply with J.

Increasing the kinetic temperature enhances the emission in lines
from the higher-energy transitions at the expense of lower-energy
transitions, similar to increases in the density, but changes in
the density are more effective because the range of reasonable
densities ($10^3 - 10^7$~cm$^{-3}$) is much larger than the range
of reasonable kinetic temperatures (10 $-$ 100K).  Because
temperature is generally less effective, a constant value of 50K
was used for the initial calculations.  Later we relaxed this
constraint because the temperatures of the interior parts of the
clouds could be several hundred degrees \citep{dn97} and higher
temperatures help the models fit the higher energy transitions,
especially the J=7$-$6 transition whose upper level has a higher
equivalent temperature than the 50K used in the initial
calculations.

\subsubsection{Effects of More than One Cloud}

When more than one cloud is present along the line of sight, the combination of the
contributions from the separate clouds can make it difficult to determine the
characteristics of the individual clouds.  For example, it is difficult to separate
the effects of turbulent velocity and the relative velocities of the clouds when two
(or more) clouds contribute to a line profile.  The solid line in Figure 1a shows
the profile of the J=3$-$2 transition of CS in the direction of H~{\sc ii} region A
that would be produced by just clouds A and G of the multi-cloud model (discussed
below in section 5.1).   The dashed line in that figure is the profile if the
turbulent velocity in each cloud is increased by 50\%, to 9.0 km~s$^{-1}$.  In
Figure 1b, the solid line is the same as before, but the dashed line now shows the
effect of changing the relative speed between the two clouds.  The turbulent
velocity remains fixed (at 6.0 km~s$^{- 1}$, same as for the solid line),
but the relative speed between the two clouds
has almost doubled (4.8 to 8.8 km~s$^{- 1}$).  The FWHM of the two dashed curves are
the same and the heights are nearly the same. The profile where the turbulent
velocities have been increased is narrower at the top and broader at the base than
the profile due to the moving clouds, but the differences are rather subtle.

When more that one cloud is present, it is sometimes useful to set
the velocities of the clouds to very different values to see what
the contribution of each cloud is to a given profile.  Figure 1c
shows the 5\arcsec, CS J=2$-$1 profile in the direction of H~{\sc
ii} region B in the preliminary multi-cloud model. In this figure
the velocity of the cloud at A has been subtracted from each of
the cloud velocities.  Figure 1d shows the effects at B of the
four clouds artificially separated in velocity that contribute
to the profile.  In this
figure the velocity of cloud A has been subtracted from each of
the clouds and then 30 km~s$^{-1}$ has been added to the cloud G',
60 km~s$^{-1}$ to G, and 90 km~s$^{-1}$ to A'.  It is apparent
from these figures that A contributes to the right part of the
profile, G to the left, G' reduces the right-hand part, and A'
produces the shoulder on the right.

\subsubsection{Limitations}

Although the tools used for this analysis assume locally spherically
symmetric flow, we can treat multiple such volumes.  However, true
3-D flow, bipolar outflows, and disk rotation cannot be modeled by the
these tools.

\section{MODELS}

\subsection{Colliding Clouds $\rightarrow$ Multi-cloud Model}

\subsubsection{Initial Model of Two Colliding Clumps}

We started with the colliding clouds model because some of the individual clouds
developed for this model are used in the other models.  This model evolved into a
multi-cloud model.  \citet{ser93} proposed a model consisting of three dense clumps.
We consider their central clump near H~{\sc ii} region G and their southwestern clump
centered on H~{\sc ii} region A.  When we tried the parameters for the clumps given
by Serabyn et al., the model line strengths were much larger than observed.  We ran a
number of cases and found a set of parameters which lead to profiles that match the
line strengths and shapes of the lines rather well except the predicted strengths of
the J=7$-$6 lines at 20\arcsec~in the directions of A and G are too weak (Figure 2a,
2b), the model does not produce the absorption observed on the high-velocity side of
the J=2$-$1 profile towards G in the 5\arcsec~BIMA observations (Fig 2d), and the
predicted strengths of the J=2$-$1 lines at 20\arcsec~in the directions of A and G
are too weak(Figure 2e).  The 7$-$6 transition will play a critical role in deciding
which models are acceptable.

\subsubsection{Two Dense Clumps and a Low-density Envelope Around One of Them}

We tried adding a low-density envelope to the uniform cloud
(clump) around H~{\sc ii} region A.  This envelope does not change
the fit to the J=2$-$1 profile at A, but it does produce
absorption at G, and the absorption is on the high-velocity side
of the profile due to the relative motion of the clouds. However,
the velocity at which the absorption appears is not high enough,
and the strength of the emission in the direction of G is too low
(Figure 3a, b). This model does produce a stronger J=7$-$6 line in
the direction of A (Figure 3c) and line shapes that match the
observed profiles of other transitions better than the models
without an envelope around A.  The increase in strength of the
J=7$-$6 line is due to line blanketing of the envelope causing
an increase in the excitation temperature for that line inside the
clump.

\subsubsection{Add More Clouds}

We then tried a separate cloud, G' in Figure 12a, behind A (so as not to cause
absorption in the profile towards A), but in front of G to produce the absorption
observed on the high velocity side of the J=2$-$1 profile at 5\arcsec~in the
direction of G. Figure 4 shows that the attempt was successful.  The profiles match
well except for the J=2$-$1 transition at 20\arcsec~and the J=7$-$6 transition.  To
increase the J=2$-$1 emission in a 20\arcsec~beam, we increased the sizes of the
clouds at A and G and reduced the molecular abundance to compensate for the
increased path length, see Figure 5. We also looked at the J=2$-$1 profile in the
direction of H~{\sc ii} region B and found that the CS emission from the clouds at A
and G fully accounted for the CS emission observed at B. The final change we made at
this stage was to add a cloud, A' in Figure 12a, at the relatively high velocity of
17 km~s$^{-1}$ behind the whole complex to account for the shoulder observed on the
high velocity side of the profile at A, see Figure 6 and Table 2. All the model
profiles match the observed ones except J=7$-$6 which are lower than observed. All
of the clouds in this model are uniform. We call this model the preliminary
multi-cloud model because, as we show below, clouds with non-uniform temperatures
and densities give profiles that match the observations better.

\subsubsection{Comparison with Other Results}

We find considerably lower H$_2$ densities than \citet{ser93}, but our [CS]/[H$_2$]
ratios are similar to those found by them and by others.  Serabyn et al. found the
molecular hydrogen density to be $6\times 10^{6}$ cm$^{-3}$ in their central clump and
$3\times 10^{6}$ cm$^{-3}$ in the southwestern clump.  We find densities a little larger than
$1\times 10^{6}$ cm$^{-3}$ for our clouds corresponding to these clumps.
Serabyn et al. use a
ratio of [CS]/[H$_2$] (per km~s$^{-1}$) of $3.6\times 10^{-10}$ for both clumps.  We
find ratios of $5.1\times 10^{-11}$ for the cloud corresponding to the central clump
and $2.6\times 10^{-10}$ for the southwestern one, and we find a turbulent velocity of
6 km~s$^{-1}$. Miyawaki, et al. (1986) estimate the [CS]/[H$_2$] ratio to be less
than or equal to $5\times 10^{-10}$.

Although the H$_2$ densities of our clumps are lower than Serabyn's values, the size
of cloud G is larger, resulting in a much larger mass for it.  Serabyn et al. find
masses for the clumps of $3\times 10^{4}$  to $6\times 10^{4}$ solar masses and a
total cloud core mass of about $1.0\times 10^{5}$ M$_{\sun}$.  Miyawaki estimated
the core mass as $.5\times 10^{5}$ M$_{\sun}$ to $2.5\times 10^{5}$ M$_{\sun}$. We
find cloud masses of $10.2\times 10^{5}$ M$_{\sun}$ and $0.8\times 10^{5}$
M$_{\sun}$ for the two main clouds with $11.3\times 10^{5}$ M$_{\sun}$ for the
total.

The discrepancy in the H$_2$ densities of the clumps will be partially resolved when
we discuss the core-envelope structure in section 5.3.  In this case, the central
H$_2$ densities are increased to $6\times 10^{6}$ cm$^{-3}$.

The V$_{lsr}$ derived for cloud G agrees with the Serabyn et al.
velocity. However, we find a V$_{lsr}$ of about 9 km~s$^{-1}$
rather than 11.5 km~s$^{-1}$ works best for cloud A with
additional, lower-density gas at V$_{lsr}$ of 13 km~s$^{-1}$ and
possibly also at 17 km~s$^{-1}$.

\subsection{Global Collapse}

We next considered a model using the parameters of a cloud in
free-fall collapse around H~{\sc ii} region G from \citet{DA94}.
We were able to reproduce our 5\arcsec~beam CS J=2$-$1
observations by using a [CS]/[H$_2$] ratio of $4.5\times
10^{-10}$. We added the cloud at A and the H~{\sc ii} region at B
from the multi-cloud model. The fit to the observed
5\arcsec~J=2$-$1 observations is satisfactory in the directions of
H~{\sc ii} regions A, B, and G; however, the model profiles are
too weak for the 20\arcsec~J=5$-$4 and 7$-$6 observations.  See
Figure 7.

\subsection{Localized Collapse}

The third set of models we considered consisted of in-falling clouds around each of
the H~{\sc ii} regions.  This model was proposed by \citet{ket91} as a result of
their analysis of the fragmentation of a rotating disk.  Our model went through
several forms before arriving at one that matched the observations.  We first
considered uniform-temperature, in-falling clouds around each of the H~{\sc ii} regions A, B,
and G.  This model was able to match the 5\arcsec~J=2$-$1 emission peaks in the
directions of A and G, but the absorption on the high-velocity side of the profile
at G was not as deep as observed, and there was too much emission in the direction
of B.  The emission at 20\arcsec~resolution was not as strong as observed in any of
the transitions.

We next considered clouds in free-fall around each of the H~{\sc ii} regions. The
model 5\arcsec~J=2$-$1 profiles matched the observed profiles rather well, but all
the 20\arcsec~profiles were much weaker than observed.  We then put a uniform cloud
around each H~{\sc ii} region inside the free-falling cloud.  All the model profiles
matched the observed ones with three exceptions. The 5\arcsec~J=2$-$1 profile in the
direction of B was a little strong and not the right shape and the 20\arcsec ~
J=2$-$1 profiles in the directions of A and G were too weak.  Increasing the size of
the cloud at A improved the fit of the 20\arcsec~J=2$-$1 profiles.  Increasing the
size of the cloud at G did not help this fit.  Adding a small, uniform, low-density
cloud in front of B, cloud B' in Figure 12b, moving at almost the speed of A
improves the fit with the observed profile at B by absorbing some of the emission in
that direction.  Physically, this material might be associated with A. The
20\arcsec~J=2$-$1 profiles are still a little weak, but the match of the model
profiles to the observed ones is good for all the rest. See Figure 8.

The local collapse model is better at matching the J=7$-$6 line than either the
multi-cloud or global collapse models.  This is most likely due to the
introduction of a core-envelope structure in cloud components for A and G.
These clouds in the local collapse model have
an 80$\%$ smaller core with a density about 5 times higher than in the
multi-cloud case and this core is surrounded by an extensive lower
density envelope.  As a result of these changes, the highest optical depth in the
J=7$-$6 line has increased by about 25$\%$ in both clouds.  The excitation
temperature of the J=7$-$6 line is about 16K throughout the
clouds for the multi-cloud model and in the envelopes for the local collapse model.
However, in the local collapse model, the excitation temperature
rises to 35K in the center of the core for cloud A and to 70K in the center
of the core for cloud G.  The overall higher optical depths
and excitation temperatures create the higher brightness temperatures
needed to fit the observed intensity of the J=7$-$6 line.

\section{MODIFICATIONS TO PRELIMINARY MODELS}

\subsection{The Preliminary Models and the J=7$-$6 Emission}

The major deficiency of the preliminary models is the weakness of
the J=7$-$6 profiles for the multi-cloud and global collapse models.
Two changes come to mind that may allow these models to reproduce
this transition: First, a higher kinetic temperature, perhaps with
a gradient, so that there would be higher temperatures in the
central regions where the density is high. Second, a core-envelope
structure which seems to have done the trick for the local
collapse case and which was hinted at already in the two-clump
model with an envelope around clump A.

\subsection{Exploration of Temperature Laws}

We know from the 20 micron emission peak just north of the H~{\sc ii} region G that
there is hot dust in the central part of W49 \citep{wt92}.  The physical conditions
in such a situation have been described in a review article \citep{ne99}.  The main
effect for our study, is that near H~{\sc ii} regions where the molecular hydrogen
density is high, the gas temperature will be comparable to the dust temperature
because of collisions. Therefore, the kinetic temperature could reach several
hundred degrees in the very center but fall off further out where the dust is less
opaque. Typically the temperature would decrease as $r^{-0.4}$ \citep{dn97, ne00}.

To see the effects on the profiles of variations in the temperature, we modified the
preliminary models as follows: 1)  changed the kinetic temperature from the constant
50K of the preliminary models to a constant 100K, 2) introduced a radial decrease in
the temperature of $r^{-0.4}$ from a high of 100K at the center, and 3) same as case
2 but with T=200K at the center. For each of these temperature structures we spread
out the cloud components in velocity space (similar to what was done in section
4.2.2 Fig. 1c) to see which components contributed most to the emission along the
different lines of sight.  For the core-envelope in-fall model, we separated the
dense, uniform, turbulent, hot core with no in-falling motions from the lower density, lower
temperature, collapsing envelope by giving each component its own offset in velocity.

\subsubsection {multi-cloud model}

As expected, increasing the temperature from 50K to 100K increases
the strength of the J=7$-$6 line so that it is a little stronger than
is observed, and the profiles for the J=2$-$1 transition are unchanged.
However, the J=5$-$4 line is now much too strong.  With a radial
gradient in the temperature from a high of 100K at the center, the
J=7$-$6 and J=5$-$4 lines are too weak.  They are weaker than when the
temperature was a constant 50K.  Increasing the central temperature
to 200K with a radial gradient causes the J=5$-$4 line to match the
observations, but the J=7$-$6 line is much too weak.

\subsubsection {global collapse model}

The effects on line profiles of changing the temperature structure
in the global collapse model are similar to those in the
multi-cloud model.  While increasing the temperature to 100K
strengthens the higher-level transitions, the J=7$-$6 line is still
too weak.  Adding a gradient in the temperature causes the J=7$-$6
and 5-4 lines to be much weaker than is observed. The J=7$-$6 line
is still too weak even if the central temperature is increased to 200K
with a gradient.  The J=5$-$4 line is about the right strength in
the direction of A, but is too weak in the direction of G.  For
all of these temperature structures, almost all of the emission in
the higher transitions comes from the cloud around H~{\sc ii}
region A.

\subsubsection {local collapse model}

Increasing the temperature from 50K to 100K causes the J=7$-$6
and J=5$-$4 lines to be much stronger than the observed lines.
Adding a gradient in the temperature reduces the strength of
these lines very nearly to what is observed.  Increasing the
central temperature to 200K with a gradient makes the J=7$-$6
and J=5$-$4 lines too strong.  They are about as strong as
when the temperature was 100K without a gradient.

\subsection{Core-envelope Models}

In section 5.1.2 we found that adding a low-density envelope to
the uniform clump around HII region A increased the strength of
the J=7$-$6 emission.  \citet{ser93} also commented on the
likihood of the clumps being immersed in lower-density gas. We now
consider clouds having a static, turbulent core with free-fall in
the surrounding volume. Adding such a core-envelope structure to
the global collapse model causes it to reproduce the J=7$-$6 and
5$-$4 transitions as well as the lower transitions.  We get the
best fit to the observations with an ``inside-out'' collapse model
\citep{shu77, shu87}. We tried three velocity structures for the
local-collapse model with the 100K temperature gradient and
core-envelope structure.  The velocity structures were: 1)
free-fall envelope with v=-5 km~s$^{-1}$ at the inner edge and
decreasing as r$^{-.5}$, 2) a constant in-fall velocity of -5
km~s$^{-1}$, 3) no in-fall velocity, which is like our multi-cloud
models but with a core-envelope structure. We find that the
differences between the profiles predicted by these models are
less than the noise in the observations.

We conclude that the core-envelope structure is the key to
reproducing the strength of the J=7$-$6 line, and to some extent the
J=5$-$4 line, while maintaining the good fits to the lower
transitions.  The temperature gradient can be adjusted to give a
good fit.  The requirement is that the highest temperatures be in
the center where the densities are sufficient.  The exact velocity
gradient in the local collapse envelopes is not critical as long as it is
in the range 0 to 5
 km~s$^{-1}$ if the turbulent velocity is around
6 km~s$^{-1}$.

\subsection{Final Models}

We arrived at our final models by adjusting the temperatures and
structures of our preliminary models.  For the multi-cloud model
we gave cloud G a core-envelope structure and made the temperature
of the core 100K.  For the global collapse and local collapse
models, we used a core-envelope structure for both clouds A and G.
The temperatures of the cores are 100K with a gradient of
$r^{-0.4}$ which continues through the envelope.  We also included
the 17 km~s$^{-1}$ cloud from the multi-cloud model behind the
other clouds of the Shu global collapse and the local collapse
models to account for the shoulder observed on the high velocity
side of the profile at A.

The physical parameters of the final models are given in Tables 3 through 6. Table 3
gives the parameters of cloud A', which is used in all three final models, and
clouds B' and B which are used in the final global collapse and local collapse
models.  Cloud A' was also used in the preliminary multi-cloud model.  In each table
the model components are listed across the top followed by their LSR velocities.
Next are the parameters of the H~{\sc ii} regions, followed by the parameters for
the components of the molecular clouds.

The profiles from the final models are displayed in Fig. 9, 10, and 11 for the
multi-cloud, Shu global collapse, and local collapse models respectively.  All three
models reproduce the observations reasonably well except for the 20\arcsec~ J=2$-$1
profiles in the directions of H~{\sc ii} regions A and G.  The model profiles are
weaker than the observed for all three models.
Figure 12 shows the clouds as they are arranged along the line of sight for the three
final models.

Let us compare the similarities and differences in the excitation temperatures
between the preliminary cloud models and the final core-envelope models with a
temperature gradient.  In both sets of models, a population inversion occurs in the lowest
transition (J=1$-$0) in the outer parts of the cloud and the excitation temperature
becomes suprathermal in the interior parts.  For the other transitions, the
excitation temperatures can be as low as a few degrees in the outer parts of
the clouds.
In the models with a temperature gradient, the kinetic temperature
is 100K in the interior and declines to around 25 to 30K in the
outer part of the envelope.  Because of the lower kinetic temperature in the outer
part of the clouds, the excitation temperatures are also lower there compared
to those for the preliminary cloud models.  However, both the J=2$-$1 and J=3$-$2 transitions  become
suprathermal in the cores with the excitation temperature exceeding 100K.
For the preliminary cloud models, only in the center of cloud G does the J=2$-$1 transition
become suprathermal ($>$ 50K).

\section{COMPARISON BETWEEN MODELS}

\subsection{Parameters}

The parameters of the H~{\sc ii} regions are essentially the same in the three
models.  The V$_{lsr}$ for cloud A is 8  km~s$^{-1}$ for the multi-cloud model and 9
km~s$^{-1}$ for the other two.  The V$_{lsr}$ for cloud G is 4.2 km~s$^{-1}$ for all
three final models.  All three models have an extra cloud A' which provides emission
on the high-velocity side of the 5\arcsec, J=2$-$1 profile in the direction of A.

The multi-cloud model has a uniform cloud for A.  Cloud G has a
core-envelope structure with higher temperature and density in the
core, but both core and envelope are uniform.  B is just an H~{\sc
ii} region, and the H~{\sc ii} region for G is separated from the
core and envelope.  Cloud G' is located behind A but in front of G
to provide absorption on the high-velocity side of the
5\arcsec~J=2$-$1 profile in the direction of G.

In the global collapse and local collapse models both clouds A and G have
core-envelope structures with H~{\sc ii} regions at their centers.  The uniform cores
have high temperature and density.  The envelopes have lower temperatures
and densities and gradients in these quantities as well as in the in-fall velocity.
B has a cloud as well as an H~{\sc ii} region.  The cloud has a high CS density and
gradients in the in-fall velocity, temperature, and density.  Cloud B' is located
behind A to absorb some of the emission in the direction of B.  These models do not
have cloud G'.  The envelope of cloud G in the global collapse model is much larger,
has a much lower density, and has a higher in-fall velocity than the envelope of
cloud G in the local collapse model.

\subsection{RMS Differences}

Table 7 gives the ratios of the rms differences between the
profiles predicted by the models and the observed profiles for the
preliminary multi-cloud model and the three final models to the
rms of the line-free baseline regions of the plots Figures 9 - 11.
The ratios show how well the theoretical profiles match the
observed compared to the observational errors.  A ratio of one
means that the rms of the theoretical profile is equal to the rms
of the observations.  Using this ratio as a figure of merit, it
appears that on the average all four models fit the observations
about equally well.  However, there are major differences in how
the errors are distributed among the transitions.  All four models
reproduce the 5\arcsec~J=2$-$1 observations very well.  However,
the fits for the 20\arcsec~J=7$-$6 transition are much better for
the three final models than for the preliminary multi-cloud model.
The poor fit for the 20\arcsec~ J=2$-$1 profiles by all three
final models has already been noted; although it appears from the
table that the multi-cloud and global collapse fits are better
than local collapse.  The fits are not very good for the
20\arcsec~J=3$-$2 transition; however they are better for the
final models than for the preliminary one.  The final multi-cloud
and global collapse models do a better job in the direction of G
and the local collapse model does better in the direction of A for
this transition. The visual impression from Figures 9-11 is that
the fits for this transition are good, and in Table 7 the size of
the rms differences for this transition are a little smaller than
for the other transitions.  The large values of the ratio are due
to the rather small value of the rms of the baseline.

\subsection{Deconvolutions}

Figures 13, 14, and 15 show the contributions of the components of the clouds that
make up the three final models to the predicted profiles.  The components are
separated in velocity as in Figure 1d.  The components are labeled and the velocity
shifts are indicated in panels c and d of each figure.  The observed profiles are
also included
 in each panel.  For each of the models, the cores of clouds A and G
produce the J=7$-$6 line.  The major differences come in the
5\arcsec~J=2$-$1 profiles in the directions of G and B as
described below.

First consider the contributions of the clouds to the emission and absorption observed
towards G.  In the multi-cloud model cloud A produces a major part of the emission with
additional contributions from A' and the core and envelope of cloud G, while G' produces
the absorption.  In the Shu global collapse model, the core of cloud G produces the
major part of the emission with contributions from A' and the core of cloud A.  The
envelope of cloud G and cloud B produce the absorption. In the local collapse model,
the cores of clouds A and G contribute equally to the emission, and cloud A' also
contributes significantly. Cloud B produces the absorption.

Now consider the contributions of the clouds to the emission and absorption
observed towards B.  In the multi-cloud model cloud A produces a
major part of the emission with additional contributions from A' and the core and
envelope of cloud G, while G' produces the absorption.  In the Shu global collapse
model, the core of cloud A is the major contributor to the emission with additional
contributions from the core of G and A'. The absorption is produced by B' with some
contribution from B and the envelope of G. In the local collapse model, the core of
cloud A is the major contributor to the emission with help from G core and A'. Cloud
B' does the absorbing with help from cloud B.

\subsection{Position-Velocity Plots}

Position-velocity plots with the continuum subtracted for the
three final models are shown in Figure 16.  The left-hand panels
correspond to cut a of \citet{miy94} and the right-hand panels to
their cut c.  Our plots are for the J=2$-$1 transition at
5\arcsec~resolution.  They also correspond to figures 8a and 8c of
paper 1 which were made with our BIMA observations along the same
cuts at the original resolution of 4.6\arcsec$\times$3.8\arcsec.
H~{\sc ii} G is close to (about 1.5\arcsec~west and 1\arcsec~north
of) the intersection of cuts a and c. The H~{\sc ii} regions are
located within their respective clouds except in the mc model
where the center of cloud G (Serabyn's central cloud) is offset
3.6\arcsec~to east and 7.4\arcsec~to the north of H~{\sc ii}
region G.

The contours in Figure 16 are intensities, I, in cgs units whereas the
corresponding plots in Figure 8 of paper 1 are given in
flux densities per beam, S\footnote{The units given in Figures 6 through 9
in paper 1 are in error;  the correct units are Jy~beam$^{-1}$~km~s$^{-1}$
in Figure 6 (same as in Figure 5) but Jy~beam$^{-1}$ ~
(not Jy~beam$^{-1}$~km~s$^{-1}$) in Figures 7 to 9.}.
The conversion between S and I may be written as follows:
\begin{equation}
   S(Jy~beam^{-1}) = \Omega(sr)~ 10^{23}~ I(cgs)
\end{equation}
where $\Omega$(sr) is the solid angle of the beam in steradians.
For a gaussian beam with half-power widths, $\theta_1$($\arcsec$) and
$\theta_2$($\arcsec$), $\Omega$(sr) may be written as
\begin{equation}
   \Omega(sr) = 1.133~ \theta_1(\arcsec)~\theta_2(\arcsec)/(42.545~10^{9})
= 10^{-9} \times [0.0266 ~\theta_1(\arcsec)~\theta_2(\arcsec)]
= 10^{-9} \times F(\theta_1(\arcsec),\theta_2(\arcsec))
\end{equation}
Substituting this expression for $\Omega$(sr) into equation 2, we obtain
\begin{equation}
   S(Jy~beam^{-1}) = 10^{14}~ I(cgs)~ F(\theta_1(\arcsec),\theta_2(\arcsec))
\end{equation}
For the CS(J=2-1) data with the original beam, F(4.6$\arcsec$ $\times$
3.8$\arcsec$) = 0.466 and for the convolved beam, F(5$\arcsec$ $\times$ 5$\arcsec$)
= 0.666.

The plots from all three models
show the same features: strong emission near H~{\sc ii} region A
with a peak at about 8 km~s$^{-1}$, absorption between 10 and 20
km~s$^{-1}$ near H~{\sc ii} region G (located close to offset 0),
the emission tends to extend
in the direction of the cloud G (the central cloud of Serabyn et
al.) with the velocity of the peak dropping to about 4
km~s$^{-1}$.  The global collapse model contours extend to greater
negative velocities (about $-15$ km~s$^{-1}$) than the other
models (about $-5$ km~s$^{-1}$).

The plots agree in general with the plots made from the
observations.  None of the plots from the models show the
northeastern cloud of Serabyn et al. because that cloud was not
included in any of our models.  All three models show less
absorption on cuts a and c than the observations do.  There are
more negative contours in our plots than in paper 1 because our
plots use a smaller interval between contours.  The emission peak
on cut a due to H~{\sc ii} region A has about the same strength,
velocity, and position as in the plot of the observations in paper
1.  However, for cut c the strongest maximum is further north
along the cut than the intersection of cuts a and c whereas in the
observations the peak is south of the intersection. In the models
the north peak is between 5 and 10 km~s$^{-1}$ whereas in the
observations it is at 10 km~s$^{-1}$. Since the P$-$V plots ``bow
out'' more to negative velocity for the global collapse model than
either of the other two models in agreement with the observations,
the P$-$V plots slightly favor the global collapse model.

\subsection{Predicted Profiles for C$^{34}$S}

We used our three final models to predict what would be observed using transitions
J=3$-$2, 5$-$4, and 7$-$6 of the C$^{34}$S molecule, and compared these predictions
to the observations reported in paper 1 and by Serabyn et al. (1993).
The comparisons are shown in
Figure 17 for the J=3$-$2 transition, Figure 18 for J=5$-$ 4, and Figure 19 for
J=7$-$6.  We first tried constructing models by using C$^{34}$S molecular data and
decreasing the CS abundance by a factor of 22 \citep{wil94}, but we found that a
factor of 15 for clouds A and A' and a factor of 13 for the other clouds gave better
fits to the observed profiles. The strengths of the predicted lines match the
observed lines in all cases except the J=7$-$6 transition towards A in the
multi-cloud model.  The weakness of the predicted line may be because this model is
the only one in which cloud A does not have the core-envelope structure which we
have found is important to produce strong J=7$-$6 lines.  We did not give cloud A
this structure because the C$^{32}$S J=7$-$6 line from this model was only a little
weaker than the observed line. In all cases the model profiles are symmetrical while
some of the observed profiles show asymmetrical or double-peaked lines. These
variations in the observed profiles may be due to noise, or they may indicate that
the models need to include additional processes such as outflows, rotation, and
clumps within an overall collapsing cloud.

\subsection{The Intensity of CS Emission away from H~{\sc ii} Regions}

Figure 6 of paper 1 shows that the integrated intensity of the CS J=2$-$1 emission
is elongated along a major axis which runs from the northeast to the southwest.
The overall width of the CS emission perpendicular to the major axis at the 25$\%$ level
is about 4\arcmin~and the width of the CS emission for the southwestern clump near H~{\sc ii}
region A at the 50$\%$ level is 2\arcmin.  Except for the very low-density envelope of
cloud G in the global collapse model which extends to 3\arcmin, the overlapping clouds in
our various models extend to 1\arcmin~or less.  Therefore, it is not surprising that predicted
profiles for the J=2$-$1 transition (with 5\arcsec~beam) for positions away from G and A
fall short of the observed strength of the CS emission.  Generally, lines-of-sight
beyond the edge of a core but within the envelope of the nearest model cloud, A or G,
are located within the 25$\%$ to 50$\%$ contours of the observed integrated intensity.
The  intensities of the predicted J=2$-$1 CS profiles at such positions are 40$\%$ to
50$\%$  of the observed values.
An exception is towards H~{\sc ii} H whose projected position is just within the core
of the G cloud and whose predicted intensity is 60$\%$ of the observed intensity.

We are obviously missing some components in our modeling such as any molecular
gas around other H~{\sc ii} regions (besides A, B, G), Serabyn et al.'s northeastern clump,
and perhaps an extensive region of low density gas into which the envelopes of the clouds merge.

\subsection{High Resolution Observations}

\subsubsection{CS J=1$-$0}

We used the three final models to predict the profiles that would be observed in the
directions of H~{\sc ii} regions A and G in the J=1$-$0 transition at resolutions of
2\arcsec~and 5\arcsec.  The 2\arcsec~profiles are shown in Figure 20 and are
deconvolved in Figure 21.  The 5\arcsec~profiles show the same features as the
2\arcsec~ones.  The main difference between the predictions of the three models is
the depth and width of the absorption in the direction of H~{\sc ii} region G.  The
absorption is much wider and deeper for the Shu global collapse model than for the
other two.  The absorption for the multi-cloud model is about 50\% deeper than for
the local collapse model.  There are subtle differences in the shape of the profile
in the direction of H~{\sc ii} region A.  All the models predict a peak about 17
km~s$^{-1}$.  Figure 21 indicates that this peak is due to cloud A'. The profiles
also show a peak about 10 km~s$^{-1}$ which is due to cloud A in the multi-cloud
model and due to cloud B' in the other two models.  The predicted strength of this
peak is greater than the one at 17 km~s$^{-1}$ for the Shu global collapse and local
collapse models, but the peaks are about the same strength for the multi-cloud
model.

\subsubsection{CS J=5$-$4}

We also used the models to predict the profiles in the J=5$-$4 transition of CS at
2\arcsec~and 5\arcsec~resolution. The 2\arcsec~profiles are shown in Figure 22, and
the profiles at 5\arcsec~are similar. At either resolution the peak in the direction
of H~{\sc ii} region A for the multi-cloud model is only about 60\% as high as for
the other two models.  The shapes of the profiles predicted by all three models are
symmetrical and are nearly the same.  Cloud A' does not contribute significantly to
any of the profiles.

\section{APPLICATION TO OTHER MOLECULES}

Radiative transfer modeling of several transitions of CS did not
allow us to clearly distinguish between the three models:
multiple, static clouds; global collapse; or local collapse.
Parameters could be found for each of these models such that the
theoretical profiles gave a reasonable match to the observed ones.
Following  the referee's suggestion, we applied the final cloud
models to other molecules, namely,  HCO$^+$ and C$^{18}$O, to see
whether either molecule could be used to discriminate between the
different models.  The observations of HCO$^+$ are from
\citep{DA94}, and the observations of C$^{18}$O are from Dickel et
al. 1999.

In applying the CS modeling results to these other molecules, the
only parameter allowed to vary is the relative abundance of the
target molecule;  all other parameters remain fixed as given in
Tables 3 to 6.

\subsection{HCO$^+$}

We first applied the radiative transfer calculations to HCO$^+$
because the initial conditions for the global collapse model for
CS came from the earlier modeling of HCO$^+$ \citep{DA94}. Aside
from the different relative abundance for the two molecules, the
main differences between the initial and final values of the cloud
parameters in the CS modeling were the higher H$_2$ density in the
core and the lower H$_2$  density in the envelope of cloud G, the
higher kinetic temperatures in its core, and the additional cloud
around H~{\sc ii} region A.

For these new calculations, we used the parameters of the clouds
determined from CS and an initial relative abundance for HCO$^+$
taken from  the paper by Dickel \& Auer. Once a  relative
abundance was chosen, that value was fixed -- both between clouds
and throughout  a given cloud.  We varied this value until we got
the best compromise in matching the theoretical  and observed
profiles  for HCO$^+$ for the J=1$-$0 transition at
7\arcsec~resolution  towards the H~{\sc ii} regions A and G and
for the J=3$-$2 transition at 24\arcsec~resolution towards the
H~{\sc ii} region G. The relative abundance of  [HCO$^+$]/[H$_2$]
is low -- about 10$^{-13~to -12}$.

Unfortunately, the predicted intensity for the J=3$-$2 transition
was too high while that of the J=1$-$0 transition was too low for
all of the relative abundances considered.  One way to remedy this
situation would be to lower the H$_2$ density in the core and
raise it in the envelope of G, i.e. closer to what was used in the
modeling by Dickel \& Auer but that was not allowed in this
exercise. Therefore, instead, we decreased n(HCO$^+$) in the high
density core and increased it in the low density envelope, but
this change was inadequate as shown for best fit obtained for
HCO$^+$ for the two collapse models in Figure 23 and the
multi-cloud model in Figure 24 (left side).

We next considered the possibility of an error in the adopted
dipole moment for HCO$^+$, $\mu$ =3.9 D  \citep{B89}.  Earlier ab
initio calculations by \citet{W75}  gave a range between 3.3 Dand
4.3 D.   For optically thin lines, an increase in $\mu $ from 3.9
D to 4.3 D or a factor, f, of 1.1 would increase the Einstein A
transition probabilities, optical depths, and critical densities
by f$^2$ =1.2.   Such an increase in the  critical density for the
J=3$-$2 transition  would make it more difficult to excite and
thus might lower its intensity  while having little effect on the
lower transitions.  However, there was no perceptible difference
between the  theoretical profiles calculated using $\mu $ = 4.3 D
and  those run with $\mu $ = 3.9 D.    Lowering $\mu $ to 3.0 D
lowered the intensities a little for both transitions.   We
conclude that changing  the value of the dipole moment within
reasonable limits does not improve the theoretical fits to the
HCO$^+$ profiles.

At this point we considered what other differences between HCO$^+$
and CS might explain the poorer fit to the HCO$^+$ observations
compared to the results of Dickel \& Auer -- for example,
differences  in excitation and/or where the molecules are located
spatially.   One such difference would be the presence of a
bipolar outflow because the abundance and hence intensity of
HCO$^+$ is often enhanced therein relative to CS. Indications that
bipolar outflows may be associated with H~{\sc ii} regions A  and
G1\&G2  in the W~49~A~North region  are the fact that  their
continuum spectra are inverted and their recombination lines are
very broad \citep{dwgwm00} .

To investigate the effect of such an outflow on the resultant
HCO$^+$ profiles, we added an outflow behind most of the other
cloud components. Because our code is not fully three-dimensional,
we attempted three approximations to the outflow: 1. an expanding
molecular shell, 2. several ``bullet'' clouds along the line of
sight towards G with velocities mimicking an outflow, and 3. a
similar series of bullets centered between H~{\sc ii} regions A
and G and extending to both A and G but with the outflow at an
angle to the line of sight.  However, for any of these to be
successful, it is necessary to lower n(HCO$^+$) in the other cloud
components by another factor of three.  All these approximations
to an outflow yield similar results.   After expanding the
calculations to C$^{18}$O (next section), we further modified the
properties of outflow case 3. Figure  24 (right side) shows the
resulting improvement in the match in the multi-cloud model when
such an outflow is added. The results for the two collapse models
are nearly identical to those of the multi-cloud model due to the
fact that HCO$^+$ is so depleted in the clouds ([HCO$^+$]/[H$_2$]
$\le$ 10$^{-13~to -12}$) that the outflow dominates in producing
the emission. We conclude that HCO$^+$ is no better than CS in
distinguishing between multiple, static clouds and either of the
collapse models.

The flexibility and ease of use of the mc program was important in
being able to create a model outflow using small clouds as
building blocks and varying their location both along the line of
sight and in the plane of the sky.

\subsection{C$^{18}$O}

We applied the same procedure to C$^{18}$O and tried to fit  the
line profile of J=2$-$1 observed with 12\arcsec~resolution by
\citet{paper1}  toward  H~{\sc ii} regions A  and G. The
intensities and line widths are reproduced with a relative
abundance, [C$^{18}$O][/H$_2$] of about $3\times$10$^{-9~to -8}$
which is equivalent to [CO]/[H$_2$] $\approx$  10$^{-5}$ and
[CO]/[C$^{18}$O] between 300 and 3000.    The results are shown in
Figure 25 for the three models -- multi-cloud, global collapse,
and local collapse.  Although the overall fit is good, the details
of the line shape are not reproduced.

The emission from CO is prominent in outflows so we assume that
C$^{18}$O is also present there.  By reducing the relative
abundance of C$^{18}$O by two-thirds in clouds A and G and then
adding an outflow (above case 3 for HCO$^+$), we end up with a
surprisingly good match to the detailed shape of the C$^{18}$O
profiles seen towards both  H~{\sc ii} regions A  and G as shown
in Figure 26 for all three models.  The only exception is towards
G in the global collapse case. These results strongly suggest that
some kind of outflow is present in the region.

\subsection{Nature of the possible outflow}

The concocted outflow is  only intended to be an approximation;
more accurate modeling is obviously warranted.  Nonetheless, this
simple approximation both improves the HCO$^+$ fits and gives very
good matches to the C$^{18}$O profiles. For this outflow, we used
four blobs, and  each blob is about one-half the size of the core
of cloud G.  The other parameters of the blobs are similar to
those found for the core of cloud G in the HCO$^+$ paper , e.g.
the H$_2$ density is closer to 10$^5$~cm$^{-3}$ rather than the
5$\times$10$^6$~cm$^{-3}$  found in the CS modeling.   There are
no velocity gradients within a blob but rather the outflow is
mimicked by spreading the blobs between H~{\sc ii} regions A  and
G with the least positive velocity (+3 km~s$^{-1}$) just east of G
and most positive  (+13 km~s$^{-1}$) towards A in the case of
C$^{18}$O.  The best parameters of the outflow for HCO$^+$ differ
slightly from those for the best outflow for C$^{18}$O in that
HCO$^+$ requires a bit more turbulence in the blobs, and the
outflow is shifted spatially and in velocity.  There is additional
HCO$^+$ in the outflow at  +1 km~s$^{-1}$ somewhat further to the
east of G and very little HCO$^+$ in the outflow at +13
km~s$^{-1}$ in the direction of  A.

For HCO$^+$ the outflow emission dominates that from the other
cloud components; whereas for C$^{18}$O the outflow plays a less
prominent role.  For C$^{18}$O the outflow may be the cause of the
asymmetry in the line profiles as it  explains the shift in
velocity of the peak of the emission profile between the
directions of the H~{\sc ii} regions A  and G.

\section{ADDITIONAL STRUCTURAL INFORMATION}

For our radiative transfer modeling of W49A North, we concentrated on three
competing dynamical models and modified the initial parameters
of the constituent clouds to get the best fit to the CS observations.
Since this investigation
began, new observations have become available, mainly concerning the embedded
ultracompact H~{\sc ii} regions which 1) show that the ring of
H~{\sc ii} regions is not
rotating, and 2) tend to favor a picture of dense clumps embedded in
an overall collapsing cloud. The evidence is as follows:

VLA observations of NH${_3}$ emission and absorption by
\citet{jk94} and our BIMA observations of CS (J=2$-$1)
\citep{paper1}, both at 4\arcsec~ resolution, show the ``C'' shape
in the position-velocity diagrams which is indicative of infalling
motions. A hydrogen recombination study by \citet{dmg97} indicates
that the H~{\sc ii} systemic velocities fall into several
groupings, similar to the molecular clumps of \citet{ser93}.  One
of their major new findings is that the systemic velocity of
H~{\sc ii} region B is more likely to be 16.5 km~s$^{-1}$ (from
H52~$\alpha$) rather than the earlier value of 2.5 km~s$^{-1}$
(from H92~$\alpha$); ~note that we arrived at a systemic velocity
of 15 km~s$^{-1}$ for cloud B by trial and error to obtain the
best fits of CS profiles towards B.
 \citet{wdwg01} observed emission from dust
and from CH$_3$CN  at 1.3 mm with the BIMA array and found both to be
 associated with some of the H~{\sc ii} regions, in particular B,
and the western part of G.  From their analysis
 of very high resolution
(0.045\arcsec) VLA images of the ultracompact H~{\sc ii} regions,
\citet{dwgwm00} conclude that to confine them, the H~{\sc ii} regions
must be surrounded by very high density molecular gas ($\sim$10$^8$~cm$^{-3}$)
and/or by turbulent gas.  A further complexity in any modeling will
be dealing with the possibility of ionized as well as bipolar outflows
associated
with H~{\sc ii} regions whose continuum has an inverted spectrum (i.e. the flux density is
increasing rather than decreasing with frequency) and
whose recombination lines are very broad ($\geq$45 km~s$^{-1}$); the
prime candidates being H~{\sc ii} regions A, and G1 \& G2.

We tend to agree with \citet{dmg97} who concluded from the above information
that the ultracompact H~{\sc ii} regions may be associated with and embedded within
individual clumps that have fragmented within a collapsing cloud, but we must
await further observations and modeling to confirm this emerging picture.

\section{CONCLUSIONS}

We started by trying to fit the observed 5\arcsec~J=2$-$1 and 20\arcsec~J=2$-$1,
3$-$2, 5$-$4, and 7$-$6 CS profiles in the direction of W49A North using three
different models: colliding clouds, global collapse, and localized collapse.  We
modified the parameters of these models until we got the best fits we could from
each model.

It is possible to find combinations of parameters for all three models that will
reproduce the observations available to us except for the 20\arcsec~J=2$-$1 profiles
observed in the direction of H~{\sc ii} regions A and G. The final models do not predict as
much emission in these profiles as is observed.  There may be an extended region of
emitting gas that is not modelled by our spherical clouds.

The observations at our disposal do not constrain the models sufficiently to yield a
unique solution.  However, there are some features that any successful model must
have: 1) H~{\sc ii} regions at the observed locations.  2) CS in emission towards
H~{\sc ii} regions G and A. 3) CS in absorption in front of  H~{\sc ii} region G and
moving towards it at the
proper speed to produce the absorption seen on the high-velocity side of the
profile.  The absorption indicates an excitation temperature in the CS J=2$-$1
transition that is lower than the brightness temperature of the H~{\sc ii} continuum
background.  This lower excitation temperature can be achieved by low H$_2$ density
in the foreground gas.  This absorption is not in front of A.  The gas doing the
absorbing may be in a collapsing cloud centered on H~{\sc ii} region G or in a
separate cloud moving towards H~{\sc ii} region G.  4) Sufficient density and high
temperature in the molecular clouds to give the observed strengths of the higher
transitions of CS. We found that a core-envelope structure is an effective way to
get the high-density region needed to reproduce higher transitions and the regions of
lower density
and temperature needed to reproduce lower transitions.

Our models indicate that high-resolution observations of the J=1$-$0 and 5$-$4
transitions of CS may distinguish between the models.  The width and depth of the
absorption in  J=1$-$0 towards H~{\sc ii} region G should distinguish between the Shu
global collapse model and the other two.  The presence of an emission peak or bump
at 17 km~s$^{-1}$ in the profile for this transition towards A would confirm the
existence of cloud A'.  The relative strengths of the peaks at about 10 km~s$^{-1}$
and 17 km~s$^{-1}$ in the J=1$-$0 transition in the direction of A may distinguish
between the multi-cloud and local collapse model and confirm the presence of cloud
B'.  The strength of the line towards A in the J=5$-$4 transition could distinguish
between the multi-cloud model and the other two.

Our models match the intensity but not the shape of the C$^{34}$S profiles
at 20\arcsec~resolution for the J=3$-$2, 5$-$4, 7$-$6 transitions.
The apparent asymmetry in the observed line profiles gradually
shifts in velocity from the J=3$-$2 to the J=7$-6$ transition.  When higher resolution
observations of these optically thin transitions become possible with adequate
sensitivity, they may better reveal the overall density structure in the W49A North region
including whether the molecular
density is enhanced around each of the H~{\sc ii} regions.  Such information would
provide additional constraints for future modeling although one would still have
some flexibility in adjusting the relative abundance of CS and C$^{34}$S.

In expanding our modeling to include the molecules HCO$^+$ and
C$^{18}$O,
 the only parameter (relative to the CS modeling) that we allowed to vary
was the relative abundance of the target molecule.   Good fits to
the C$^{18}$O profiles were obtained, but  no values of the
HCO$^+$ abundance could be found that would fit satisfactorily
both the J=1$-$0 and J=3$-$2 lines.

While the additional modeling of HCO$^+$ and C$^{18}$O provided no
clear indication as to which of the three models is the closest to
reality, it did result in our adding a bipolar outflow which
improves the fit to the HCO$^+$ profiles and nicely explains the
changes in the asymmetries in the  C$^{18}$O profiles as a
function of position.   The poorer fit for C$^{18}$O  towards G
for the global collapse model with or without an outflow gives a
slight preference to either the multi-cloud model or the local
collapse model over the global collapse model.

The expanded modeling also highlighted the flexibility and
usefulness of the ``mc'' program (described in the appendix) in
exploring possible spatial configurations.

Observations to test the local-collapse model include high-resolution
observations of sub-millimeter dust emission to determine whether there are
individual clumps around the H~{\sc ii} regions.  At present, the results are
inconclusive; there is indirect evidence for dense molecular gas around
the H{\sc ii} regions in order to confine them, and dust emission has been
observed associated with ultracompact H~{\sc ii} regions B and G1 \& G2.
If the individual clumps around the H~{\sc ii} regions are confirmed,
then the velocity field could be probed with
high-resolution observations of a molecular line which is both a
high-density tracer and an optically thin transition. The parameters of our final
models could be used as input to a two-dimensional radiative transfer code so that
possible rotation of the system of clouds and bipolar outflows could be
realistically investigated.  The parameters we found could be used with a full
three-dimensional radiative transfer code to properly treat multiple clouds embedded
in a low-density envelope.

\acknowledgments

J. A. Williams gratefully acknowledges sabbatical leaves in 1990
and 1997. He also gratefully acknowledges the hospitality extended
by the University of Illinois Astronomy Department and the use of
BIMA computers for data reduction and modeling during those
sabbaticals and during numerous shorter visits to work on this
project. J. A. Williams and H. R. Dickel gratefully acknowledge
partial support from the Laboratory for Astronomical Imaging which
is operated with funds provided by the Berkeley-Illinois-Maryland
Association. The BIMA research was partially supported by the
National Science Foundation through grants AST 90-24603, 93-20239,
and 96-13999 to the University of Illinois. This work was also
partially supported by a grant from the Hewlett-Mellon Fund for
Faculty Development at Albion College, Albion, MI, and by the
Office of Academic Affairs at Albion College. The J=3$-$2 HCO$^+$
data were obtained with the 12 m telescope which at the
 time was operated by the National Radio Astronomy Observatory, a facility
 of the National Science Foundation operated under
 cooperative agreement by Associated Universities, Inc. The authors thank
Dr. N.~J. Evans for illuminating discussions regarding infrared
emission from star-forming regions and its possible effects on
molecular excitation.   Although it meant more work to follow
through on the anonymous referee's suggestion to model HCO$^+$ and
C$^{18}$O, the results were well worth the work and the delay in
publishing the paper.

\appendix

\section{DESCRIPTION OF THE PROGRAM MC}

Mc is a Fortran program which runs under X Windows to plot the
output intensities from the cloud models made by the rt program.
The purpose of mc is to permit the user to investigate
interactively the effect that changes in the assumed physical
structure of the region have on the predicted outgoing radiation.
Mc is based on a plotting program written by L. H. Auer.  Its
inputs are the output files of rt, namely files of emergent
intensity and optical depth as a function of impact parameter,
velocity, and transition for a given molecule.  Mc also reads a
user-created file which gives the number of clouds, the distance
of the complex, the names of files containing observed profiles,
and for each cloud in the model the names of the intensity and
optical depth files from rt and the velocity and position offset
of that cloud from the origin.  The user interactively specifies
the kind of plot: intensity versus position along a cut, intensity
versus velocity, or intensity as a function of position and
velocity.  The user also specifies the beam size, the impact
parameters for the profile or the position of the cut, the
transitions to plot, and whether or not the continuum should be
subtracted.  For line profiles the user also specifies which
observed profiles should appear in the plot.  After the plot is
produced, the user is given the option of making more plots with
the same input files.  While the program is running the user has
the option of changing the velocity and/or position offset of any
of the clouds.

The outputs consist of the plot in the X Windows Tektronics window
and in a postscript file which could be sent to a laser printer, a
log file with the information from the user-created input file and
the user's interactive inputs, and a file that contains the
intensity versus velocity profiles that were computed to create
the plot. For line-profile plots the rms difference between the
model profile and the observed profile and some parameters
describing the model profile are displayed in the X Windows VT
window and written to the log file.

\section{NUMERICAL DETAILS}

\subsection{Intensity Along a Line of Sight}

The display program, mc, computes the intensity along a given line
of sight for a given velocity and transition as
\begin{equation}
i = \sum{i_{out} + i_{in} e^{-\tau}}
\end{equation}
The sum is over the clouds in the line of sight from the most distant cloud to the
closest.  The $i_{in}$ is the intensity that has accumulated so far.  For the first
cloud it is the 3K background.  $\tau$ is the opacity of the cloud along that line of
sight for the given velocity and transition, and $i_{out}$ is the emergent intensity
given by the program rt.  The rt program correctly handles the effects of radiation
on the populations of the energy levels in any one cloud, and the mc program assumes
that the radiation from one cloud does not affect the populations of the energy
levels of other clouds.

\subsection{Justification for Ignoring the Effects of Radiation from One Cloud
on Another}

This assumption is true only when the relative velocities of the
clouds are large enough that the line profiles do not overlap.
However, even when the profiles do overlap, radiation from
locations within a particular cloud probably has more effect on
the populations of the molecular energy levels within that cloud
than radiation from an external cloud.  This larger effect is
because the solid angle subtended by the external cloud will
usually be much less than that subtended by the cloud that the
particular point is in.  One can see from Figure 27 that
\begin{equation}
\sin {\theta _0} = r_2 / (r_1 + r_2),
\end{equation}
or if $r = r_2 / r_1$, then
\begin{equation}
\theta _0 = \arcsin (r/(1+r)).
\end{equation}
Integrating over the portion of a unit sphere with polar angle
$\theta$ gives the solid angle, $\Omega$, of the external cloud.
\begin{equation}
\Omega = \int_0^{\theta _0} 2 \pi \sin{\theta} d \theta = 2 \pi (1 -
\cos{\theta _0})
\end{equation}
There are 4$\pi$ steradians in a full sphere, so the percentage of
the celestial sphere at A covered by the external cloud is just
$50(1 - \cos{\theta _0})$.  Table 8 gives some sample values.

\subsection{Convolution}

For beam widths larger than zero, a gaussian-weighted sum is
formed of the intensities at grid points on the sky within 1.5
beam widths of the desired position.  The spacing of the grid
points is one twentieth of the beam width.

\clearpage

\figcaption {Model spectra (a) The effect on a profile from two
clouds of changing only the turbulent velocity.  (b) The effect on
a profile of changing only the relative speed of two clouds in the
line of sight.  (c) The profile in the direction of H {\sc ii}
region B from the preliminary multi-cloud model.  The zero of the
velocity scale has been set to the velocity of cloud A.  (d) To
show the contributions of the individual clouds to the profile in
(c), multiples of 30 km~s$^{-1}$ have been added to the velocities
of the individual clouds.  The clouds are designated by the
letters used in Table 1 and the short vertical lines
indicate the velocity offsets. \label{fig1}}

\figcaption {Comparison of CS spectra from the final two-clump model with
observations in the directions of H~{\sc ii} regions A and G. The histograms are the
observed profiles, and the smooth curves are the model profiles.  (a) and (b) The
predicted strengths of the J=7$-$6 lines are less than observed. (c) The predicted
line towards H~{\sc ii} region A matches the BIMA observations. (d) The predicted
line towards H~{\sc ii} region G does not have the absorption on the high velocity
side of the profile that is evident in the BIMA observations.  (e) The predicted
strengths of the J=2$-$1 transition with 20\arcsec~beam in the directions of A and G
are less than observed. \label{fig2}}

\figcaption {Two-clump model with envelope around clump A.  (a)
Profile for CS J=2$-$1 in the direction of H~{\sc ii} region A
with a 5\arcsec~beam.  (b) Profile for J=2$-$1 in the direction of
H~{\sc ii} region G with a 5\arcsec~beam.  (c) Profiles for other
transitions in the direction of A with a 20\arcsec~beam.
\label{fig3}}

\figcaption {Results for the initial multi-cloud model.  The panels of the rest of
the figures showing results of models (Figs. 5$-$11 follow this same pattern of
presentation).  (a) Observed and predicted profiles for higher transitions of CS in
the direction of H~{\sc ii} region A with a 20\arcsec~beam. (b) Observed and
predicted profiles for higher transitions of CS in the direction of H~{\sc ii}
region G with a 20\arcsec~beam.  (c) Observed and predicted profiles for the J=2$-$1
transition of CS in the direction of A with a 5\arcsec~beam.  (d) Observed and
predicted profiles for the J=2$-$1 transition of CS in the direction of G with a
5\arcsec~beam.  (e) Observed and predicted profiles for the J=2$-$1 transition of CS
in the directions of A and G with a 20\arcsec~beam. \label{fig4}}

\figcaption {Results for the intermediate multi-cloud model.  The radii of the
clouds at A and G have been increased to better represent the CS J=2$-$1 transition
in a 20\arcsec~beam.  (f) Observed and predicted profiles for the J=2$-$1 transition
of CS in the direction of H~{\sc ii} region B. \label{fig5}}

\figcaption {Results for the preliminary multi-cloud model which
now includes an additional 17 km~s$^{-1}$ cloud at the rear.  The
predicted emission in the J=7$-$6 transition in the directions of
A and G is less than observed. \label{fig6}}


\figcaption {Results for the preliminary global collapse model
where H~{\sc ii} region A and a cloud surrounding it have been
added. There is not enough emission in the higher transitions
especially in the direction of G.  There is not enough emission in
the J=2$-$1 transition with a 20\arcsec~beam, especially in the
direction of G. \label{fig7}}

\figcaption {Results for the preliminary local collapse model with
uniform, static cores surrounded in the case of A and G by envelopes in
free fall.  A small, low-H$_2$ density cloud (component B', which
may be an extension of cloud A) has been added in front of H~{\sc
ii} region B.  Now the higher transitions and the emission in the
direction of B fit well, but the emission from the J=2$-$1
transition with a 20\arcsec~beam is still weak. \label{fig8}}

\figcaption {Results for the final multi-cloud model.  The
predicted emission from the J=7$-$6 transition in the direction of
A and from the J=2$-$1 transition with 20\arcsec~beam in the
direction of G are weaker than observed.  \label{fig9}}

\figcaption {Results for the final Shu global collapse model.  The
predicted emission from the J=2$-$1 transition with 20\arcsec~beam
is weaker than observed especially in the direction of G. \label
{fig10}}

\figcaption {Results for the final local collapse model.  The
predicted emission from the J=2$-$1 transition with 20\arcsec~beam
is weaker than observed. \label{fig11}}

\figcaption {Relative sizes and locations of clouds along the line
of sight. The shaded circles are H~{\sc ii} regions.  The observer
is a the bottom of each panel.  The horizontal displacement is in the E-W
(RA) direction on the plane of the sky with the scale indicated at the
bottom.
(a) Multi-cloud model. The circles
around G H~{\sc ii} and B are dashed to indicate that the CS
abundance has been set so low that these clouds are essentially
H~{\sc ii} regions.  (b) Local collapse model.  The dashed circles
in clouds A and G represent the boundaries between the inner
uniform, static cores and the outer free-falling envelopes. (c) Global
collapse model. \label{fig12}}


\figcaption {Deconvolved profiles from the final multi-cloud
model.  The plots are similar to Figure 1d except the velocity of
cloud A has not been subtracted from the velocity of each cloud
before shifting it along the velocity axis. The vertical lines
with the labels mark the location of the cloud components; the
shorter vertical line to the left of each label gives the location
of 0.0 km~s$^{-1}$ for that component (and thus indicates the
amount of the velocity shift).  The observed profiles are also
included in the plots.  Cloud A and the core of cloud G are the
major contributors to the J=7$-$6 line.  Cloud A produces a major
part of the emission in the 5\arcsec~J=2$-$1 profile in the
direction of G, and G' produces the absorption observed on the
high velocity side of that profile. \label{fig13}}

\figcaption {Deconvolved profiles from the final global collapse
model (similar to Figure 13). The cores of clouds A and G are the
major contributors to the J=7$-$6 lines. The core of cloud G
produces the major part of the emission in the 5\arcsec~J=2$-$1
profile in the direction of G, and cloud B and the envelope of
cloud G produce the absorption observed on the high velocity side
of that profile. \label{fig14}}

\figcaption {Deconvolved profiles from the final local collapse
model (similar to Figure 13).  The cores of clouds A and G are the
major contributors to the J=7$-$6 lines.  The cores of clouds A
and G contribute equally to the 5\arcsec~J=2$-$1 profile in the
direction of G.  Cloud A' also contributes significantly.  Cloud B
produces the absorption observed on the high velocity side of that
profile. \label{fig15}}


\figcaption {Position-velocity plots from the three final models
along the a and c cuts used by Miyawaki et al. (1994).  Cut a is
parallel to the main axis of the molecular cloud with values
increasing to the east; cut c is perpendicular to the main axis
with values increasing to the north.  Offsets are measured from
the intersection of cuts a and c which is close to H~{\sc ii} G.
The H~{\sc ii} regions are located within their respective clouds
except in the mc model where cloud G (Serabyn's central cloud) has
a projected offset from H~{\sc ii} region G of $\sim$
7\arcsec~along cut a and $\sim$ 7\arcsec~along cut c. The velocity
and offset position (of the projection) of a cloud component along
the two cuts are labeled with its letter. The contours are in
units of $10^{-14}$ erg s$^{-1}$ cm$^{-2}$ Hz$^{-1}$ sr$^{-1}$.
For cut a, the intensities go from -0.585 to 4.23 in steps of
0.535 which is equivalent to -0.27 to 2.0 Jy~beam$^{-1}$ in steps
of 0.25; this range is similar to that in Figure 8 of paper 1
except they used an interval of 0.5 (= 2$\sigma$) rather than
0.25.  For cut c, the intensities go from -1.16 to 2.54 in steps
of 0.37 which corresponds to -0.52 to 1.2 Jy~beam$^{-1}$ in steps
of 0.17. \label{fig16}}

\figcaption {Observed and predicted profiles for the J=3$-$2
transition of C$^{34}$S towards A and G for a 20\arcsec~beam for
the three final models.  The profiles towards H~{\sc ii} region A
are on the left and those towards G are on the right.  Profiles
from the multi-cloud model are in the first row, from the global
collapse model are in the second and from the local collapse model
are in the third.  The panels in Figures 18 $-$ 23 follow the same
arrangement. \label{fig17}}

\figcaption {Observed and predicted profiles for the J=5$-$4
transition of C$^{34}$S towards A and G for a 20\arcsec~beam for
the three final models. \label{fig18}}

\figcaption {Observed and predicted profiles for the J=7$-$6
transition of C$^{34}$S towards A and G for a 20\arcsec~beam for
the three final models. \label{fig19}}

\figcaption {Predicted profiles for the J=1$-$0 transition of CS
towards A and G with a 2\arcsec~beam for the three final models.
\label{fig20}}

\figcaption {Deconvolved profiles for the J=1$-$0 transition of CS towards A and G
with a 2\arcsec~beam for the three final models. \label{fig21}}

\figcaption {Predicted profiles for the J=5$-$4 transition of CS
towards A and G with a 2\arcsec~beam for the three final models.
\label{fig22}}


\figcaption {Observed and predicted profiles for HCO$^+$ for the
J=1$-$0 transition towards A and G for a 7\arcsec~beam and for
the J=3$-$2 transition  towards G for a 24\arcsec~beam :
 left side is for the final global collapse model and right side is
for the final local collapse model.  Note how the predicted
intensities for the J=1$-$0 transition are too weak while those
for J=3$-$2  are too strong. \label{fig23}}

\figcaption {Observed and predicted profiles for HCO$^+$ for the
final multi-cloud model for the J=1$-$0 transition towards A and G
for a 7\arcsec~beam and for  the J=3$-$2 transition  towards G for
a 24\arcsec~beam: left side without an outflow  and  right side
with the addition of an outflow. The outflow provides the needed
increase in the intensity of the J=1$-$0 emission without
exceeding the observed line strength for J=3$-$2. \label{fig24}}

\figcaption {Observed and predicted profiles for the J=2$-$1
transition of C$^{18}$O for a 12\arcsec~beam  for the three final
models:  left side towards A and right side towards G.  The
overall fit is generally good but the asymmetries at the peak of
the emission are not matched. \label{fig25}}

\figcaption {Observed and predicted profiles for the J=2$-$1
transition of C$^{18}$O for a 12\arcsec~beam  for the three final
models but with the addition of an outflow:  left side towards A
and right side towards G. The addition of an outflow allows the
asymmetric nature of the profiles to be matched better except
towards G in the global collapse model. \label{fig26}}

\figcaption {Diagram of the angle subtended by the radius of a
spherical cloud seen from a point outside the cloud.  The
percentage of the celestial sphere covered by the cloud as seen
from A is $50(1-\cos (\theta_0))$. \label{fig27}}

\newpage
\plotone{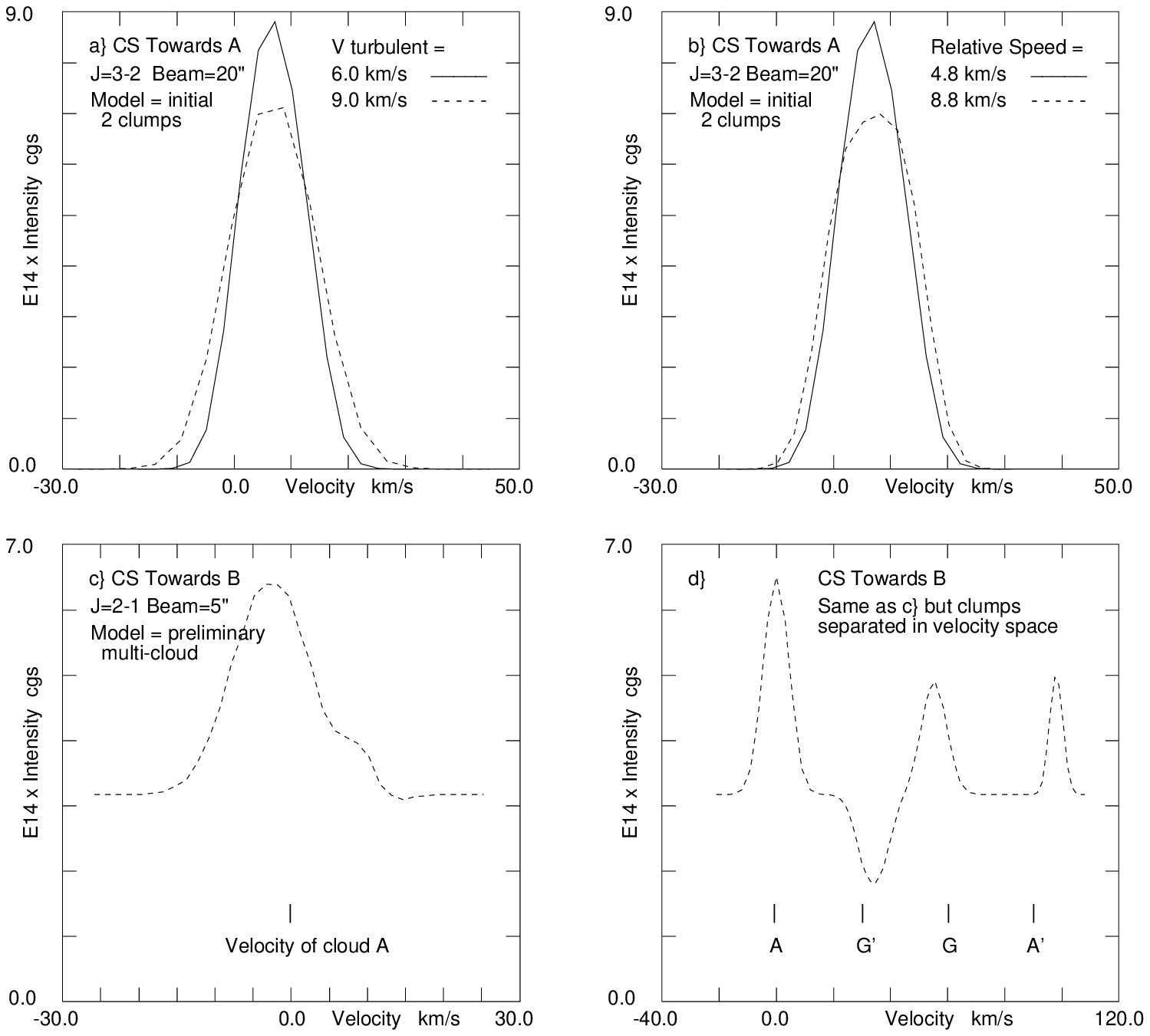}
\newpage
\plotone{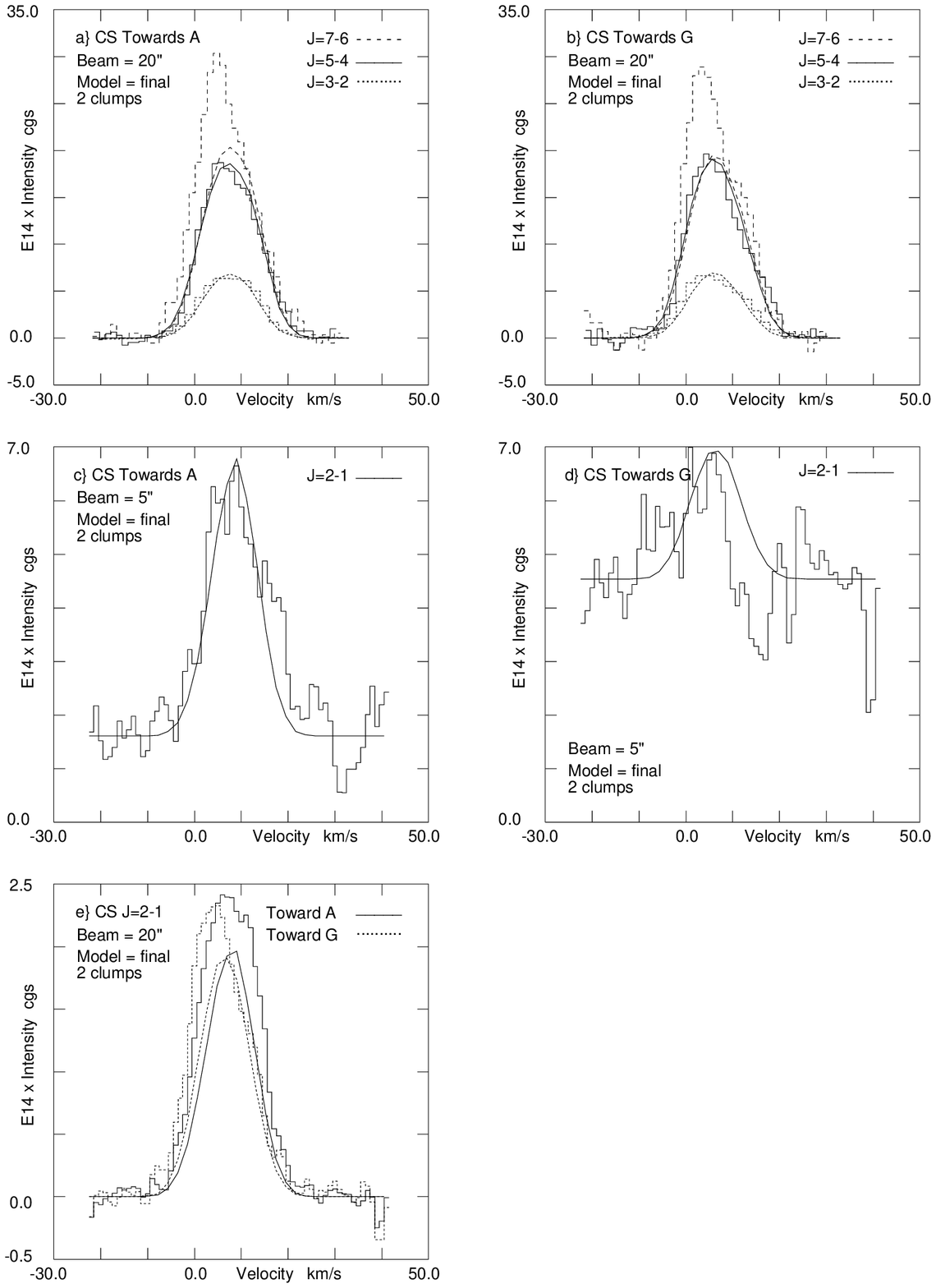}
\newpage
\plotone{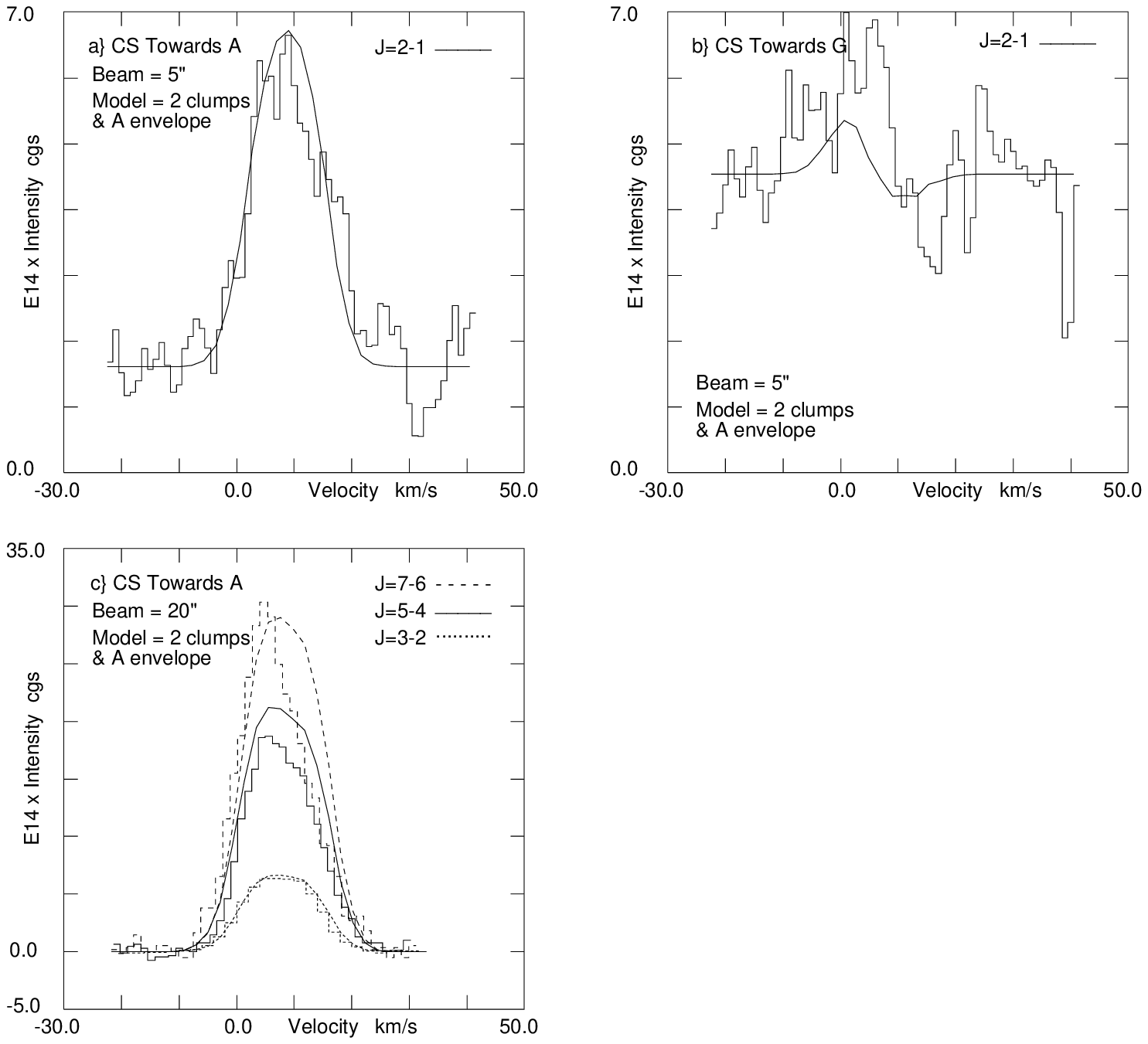}
\newpage
\plotone{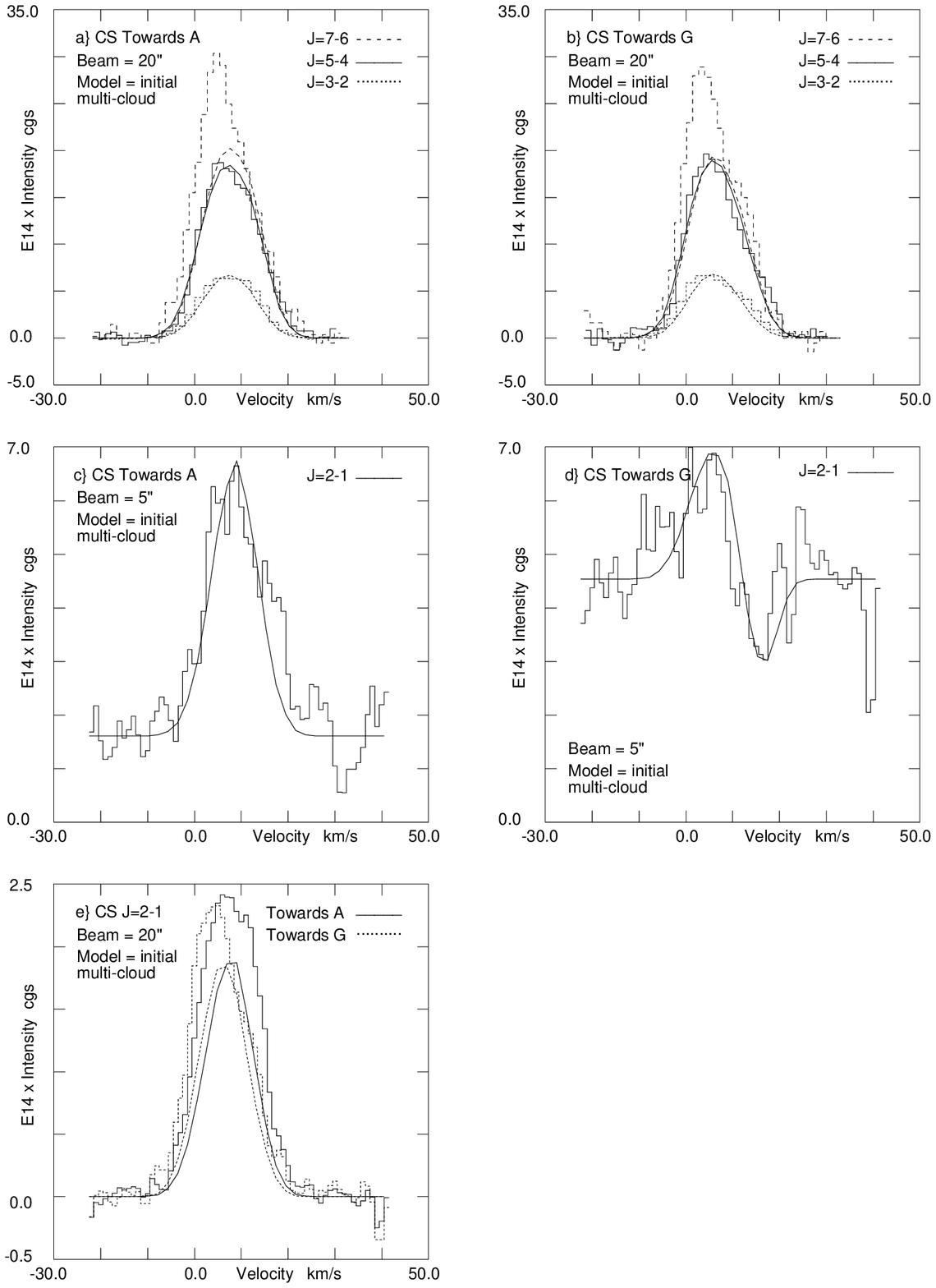}
\newpage
\plotone{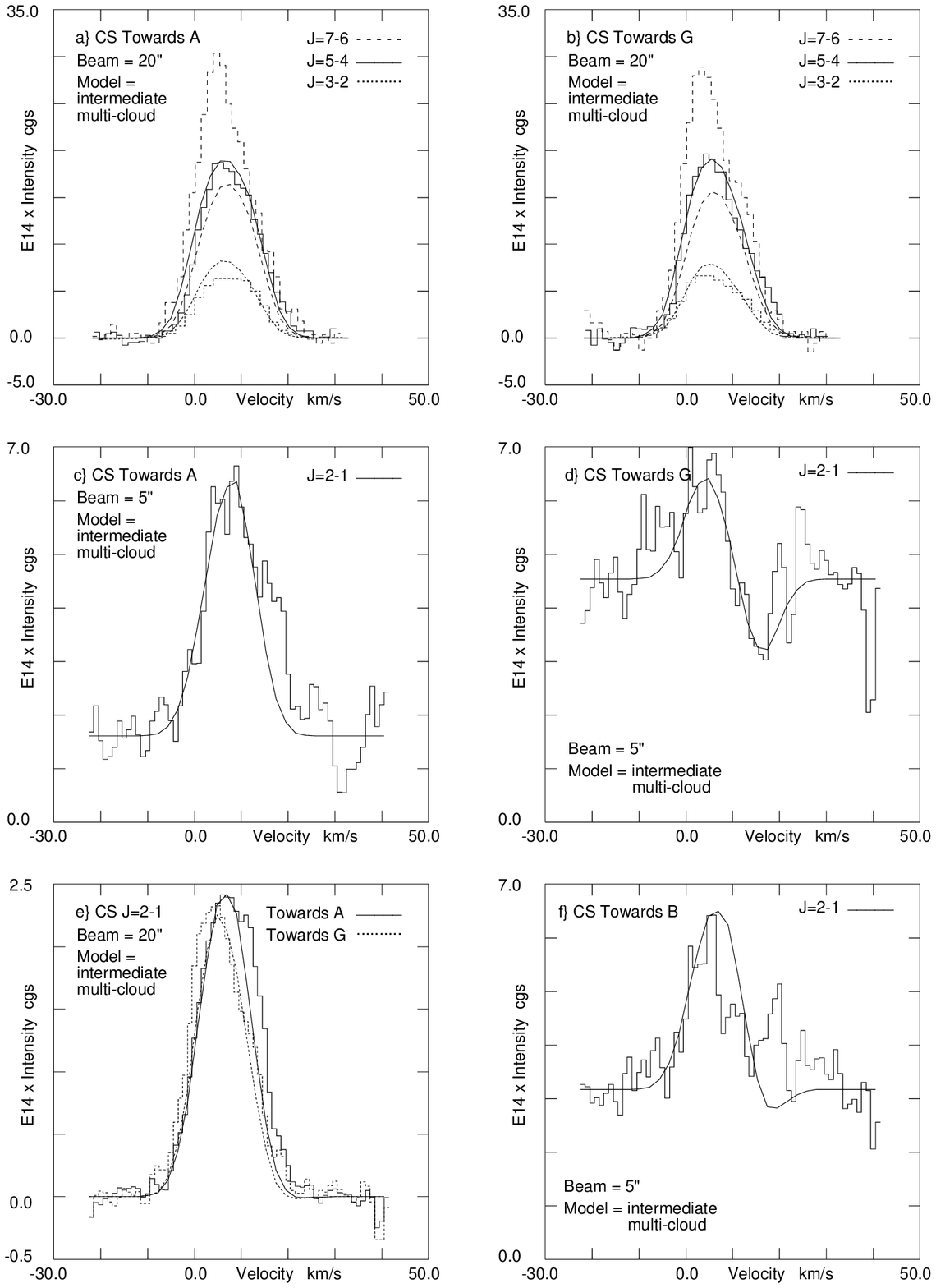}
\newpage
\plotone{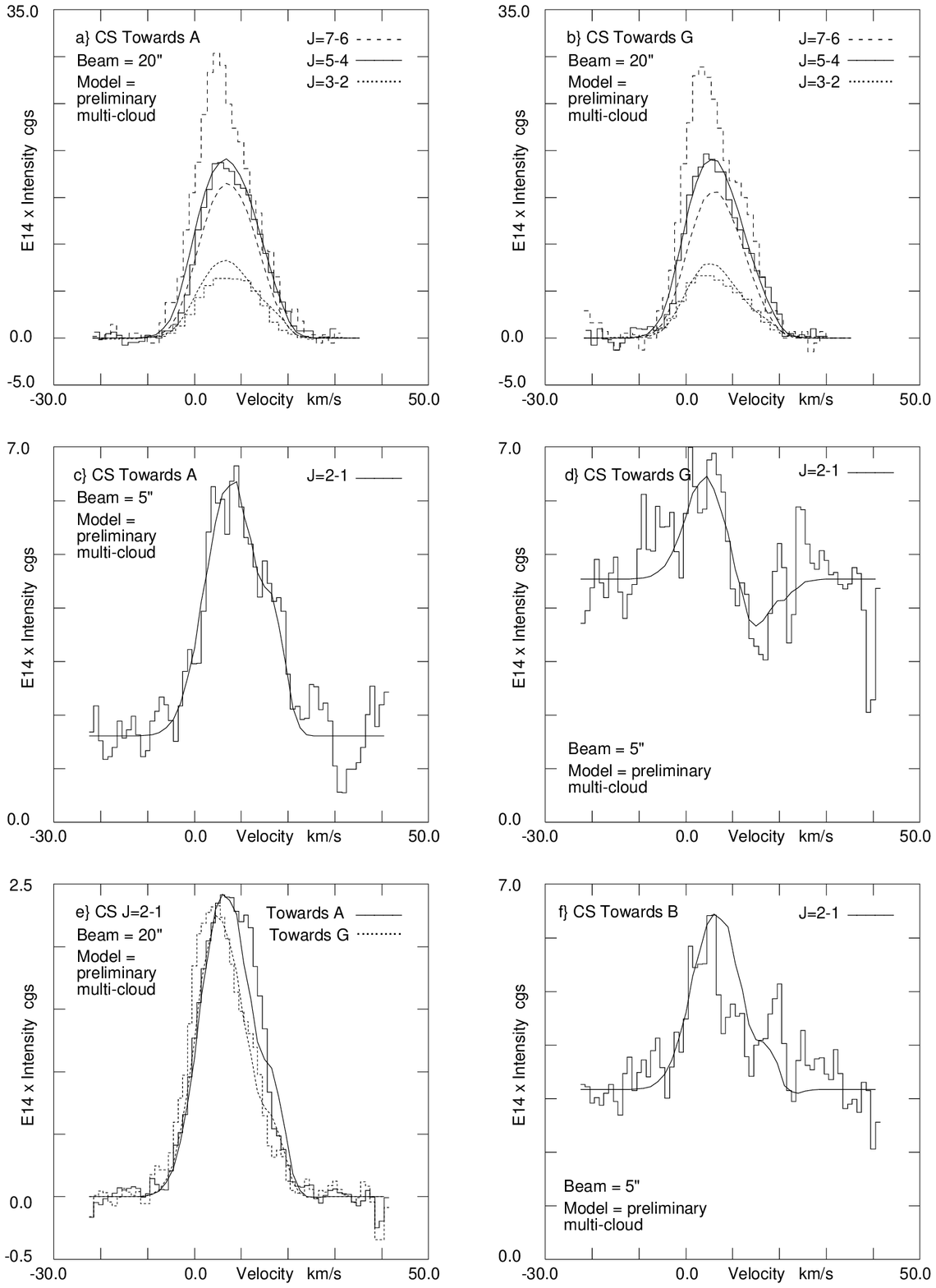}
\newpage
\plotone{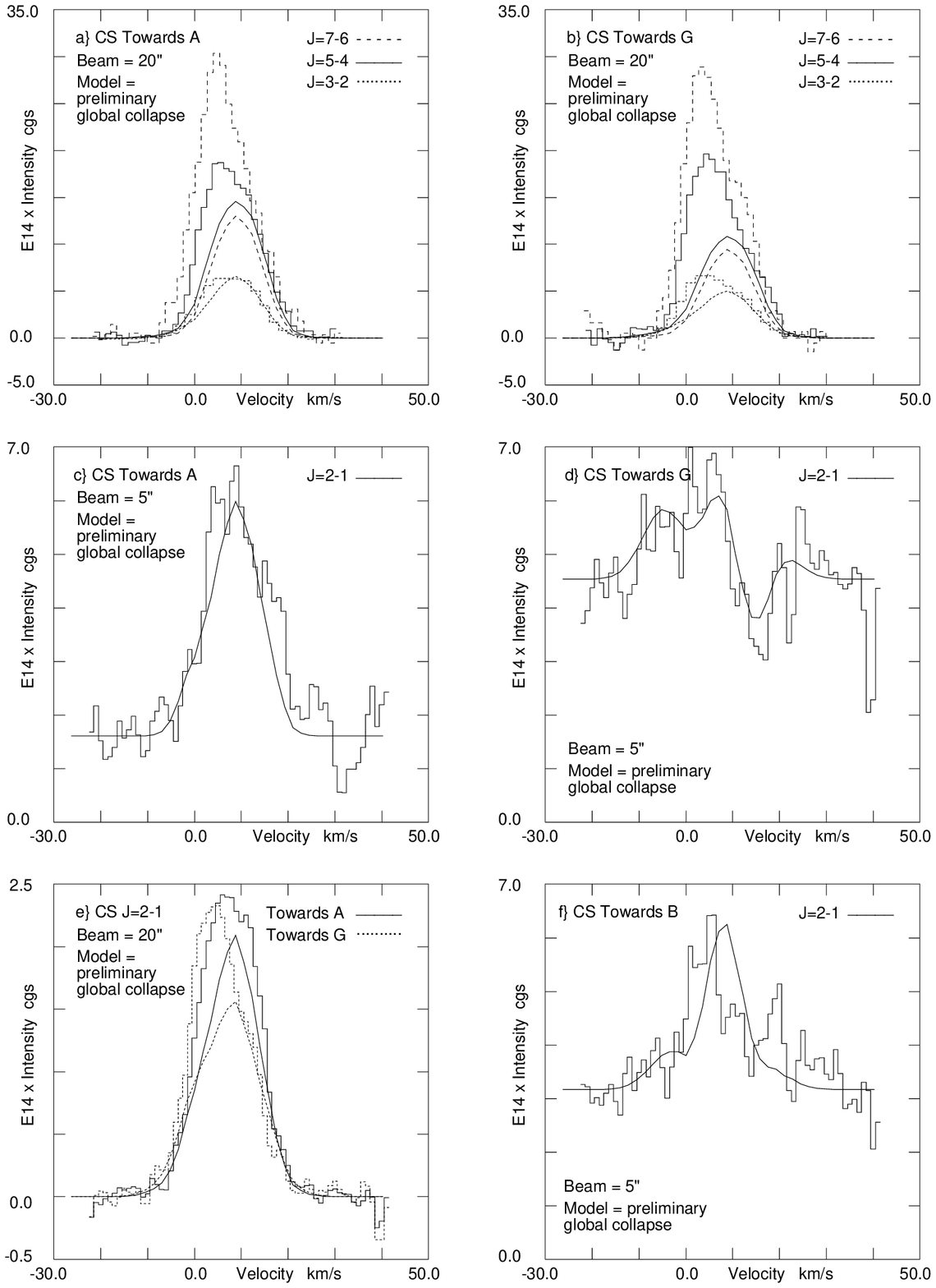}
\newpage
\plotone{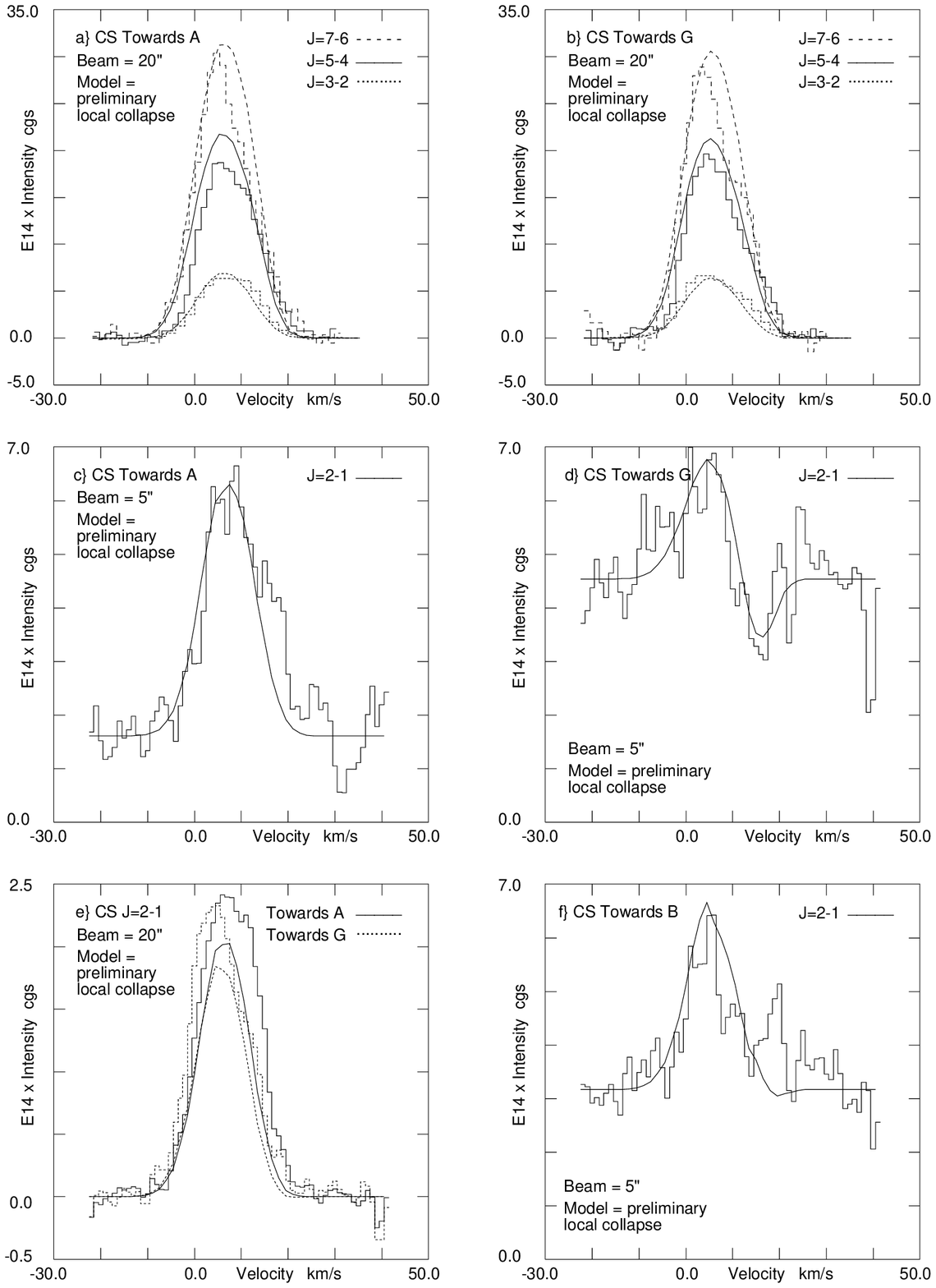}
\newpage
\plotone{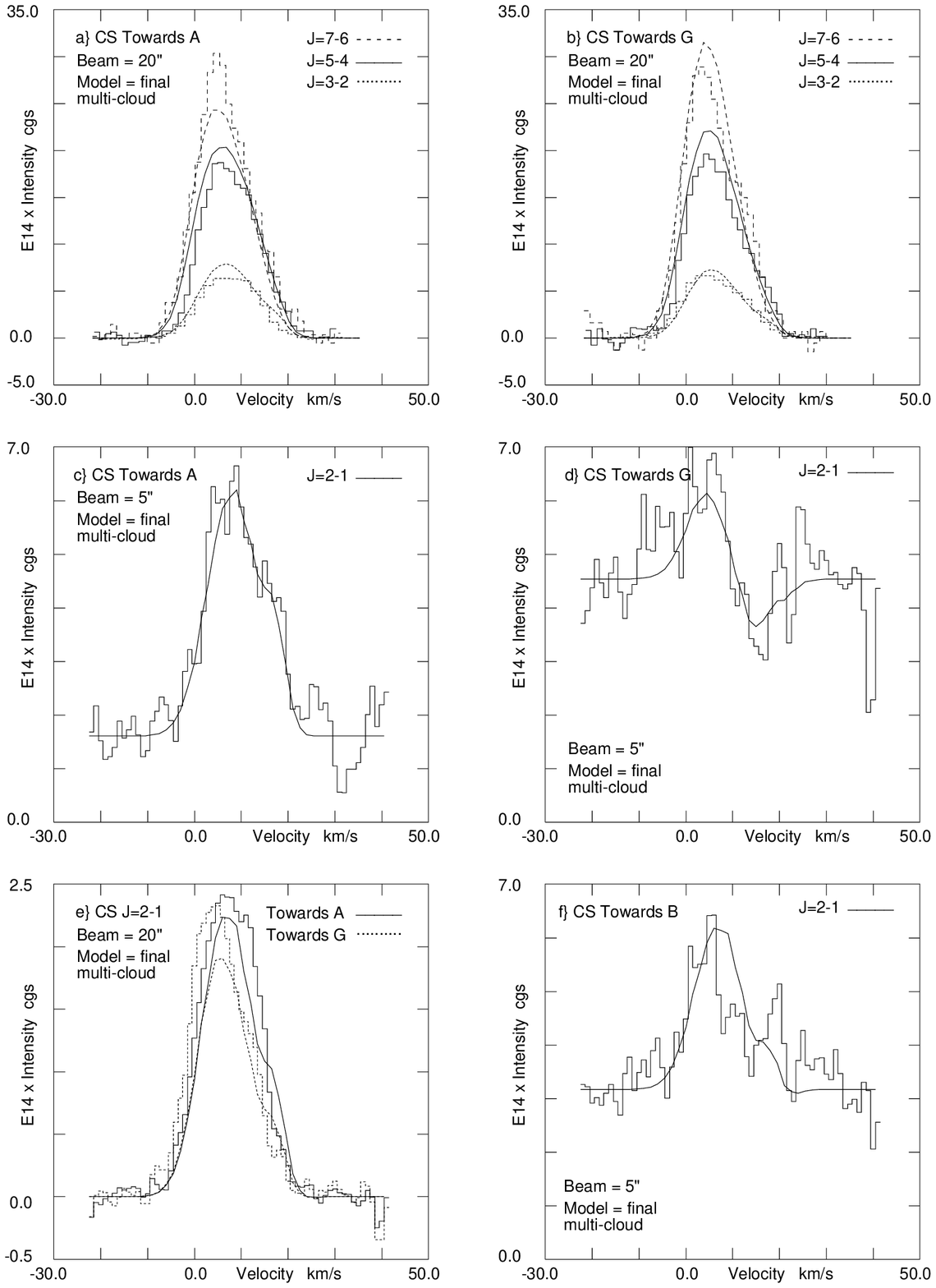}
\newpage
\plotone{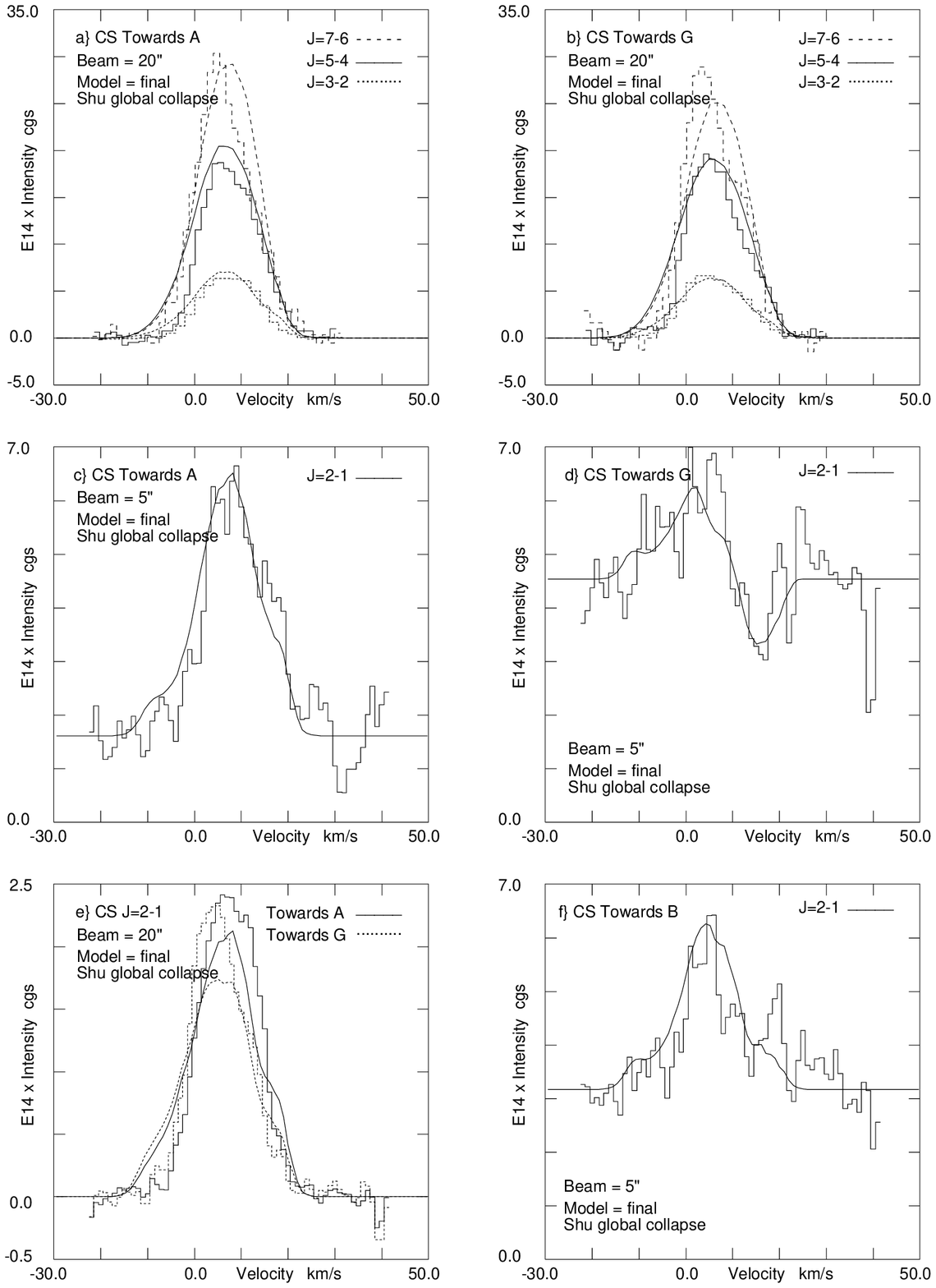}
\newpage
\plotone{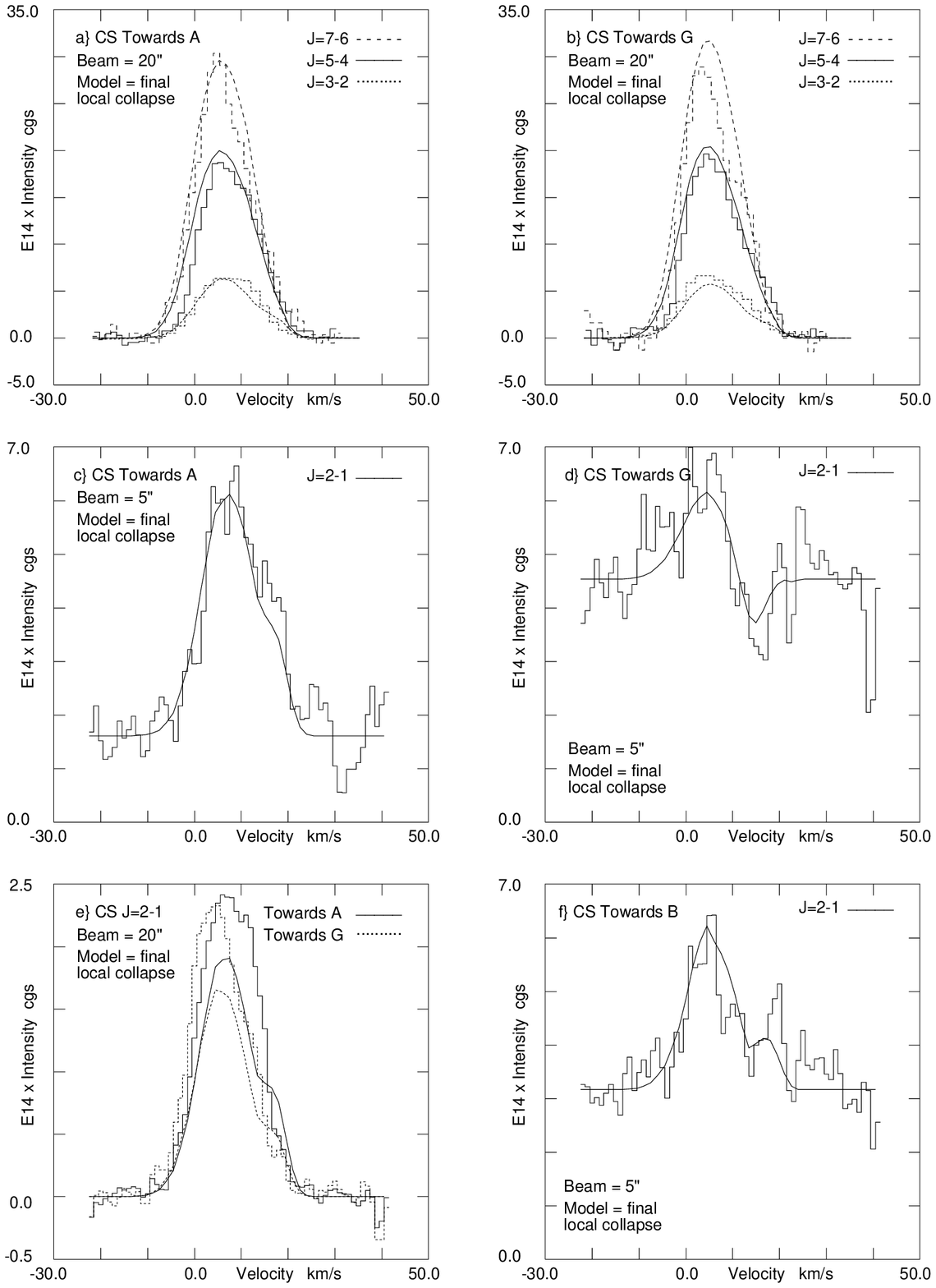}
\newpage
\epsscale{.70}
\plotone{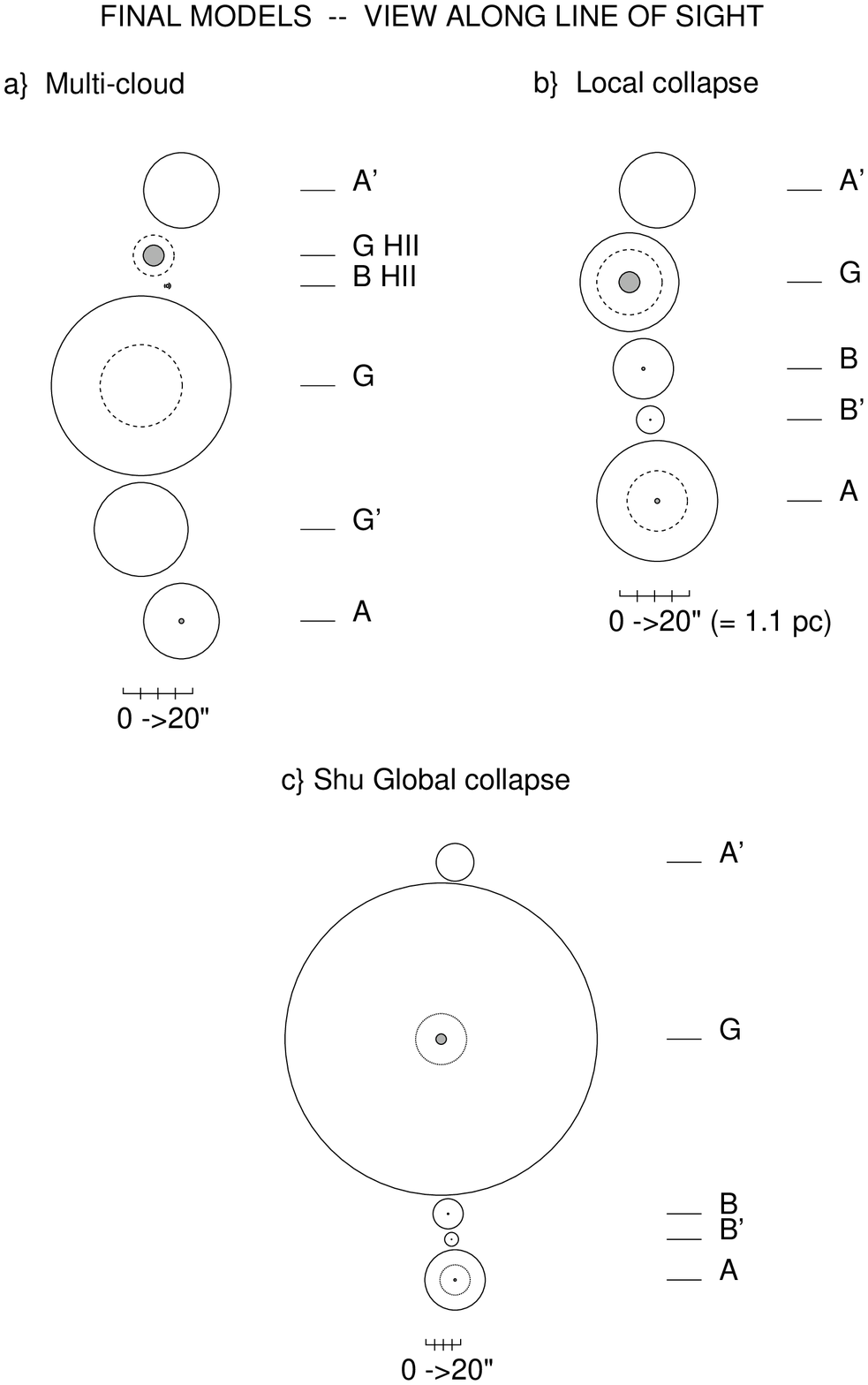}
\newpage
\epsscale{1.0}
\plotone{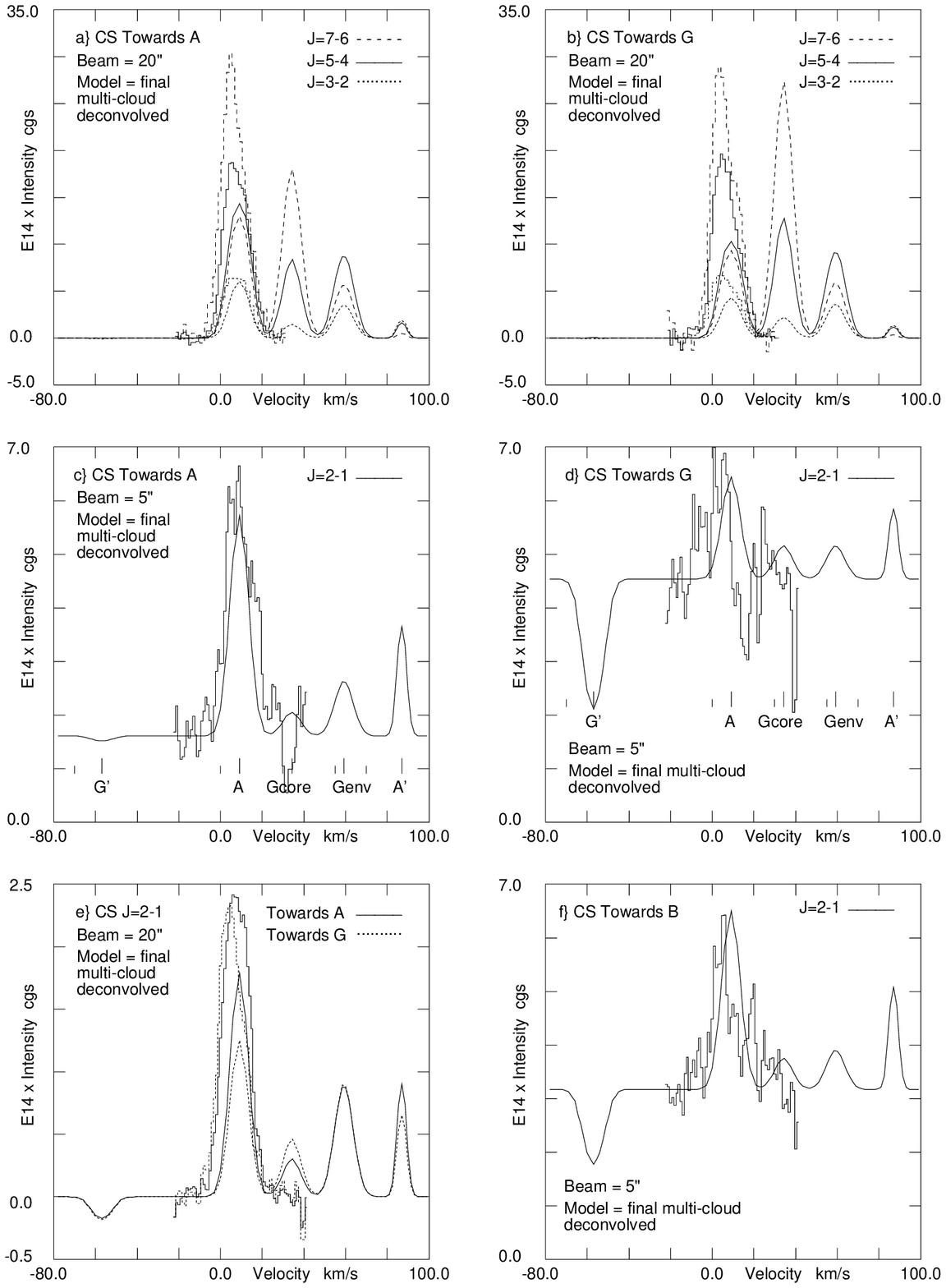}
\newpage
\plotone{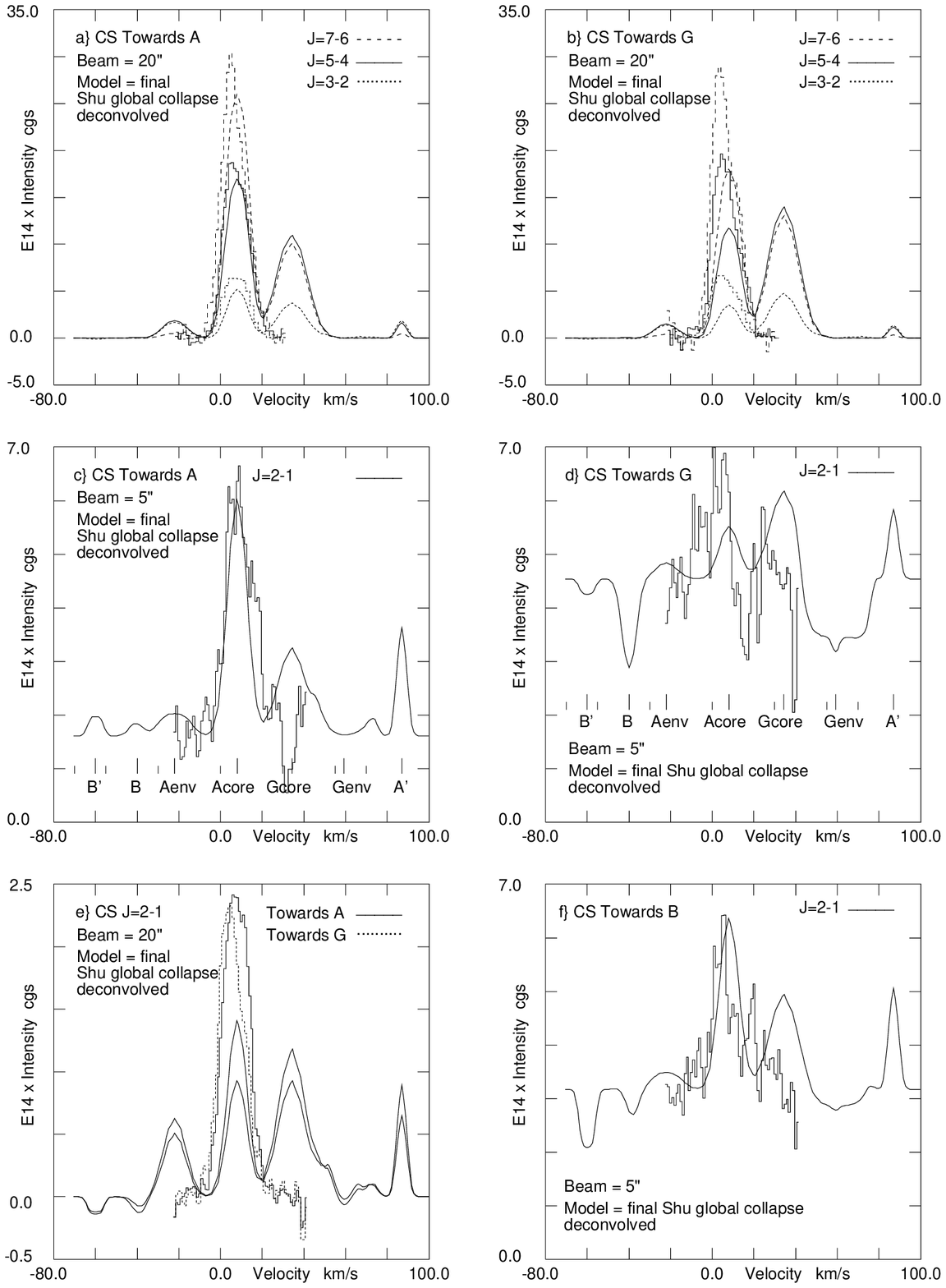}
\newpage
\plotone{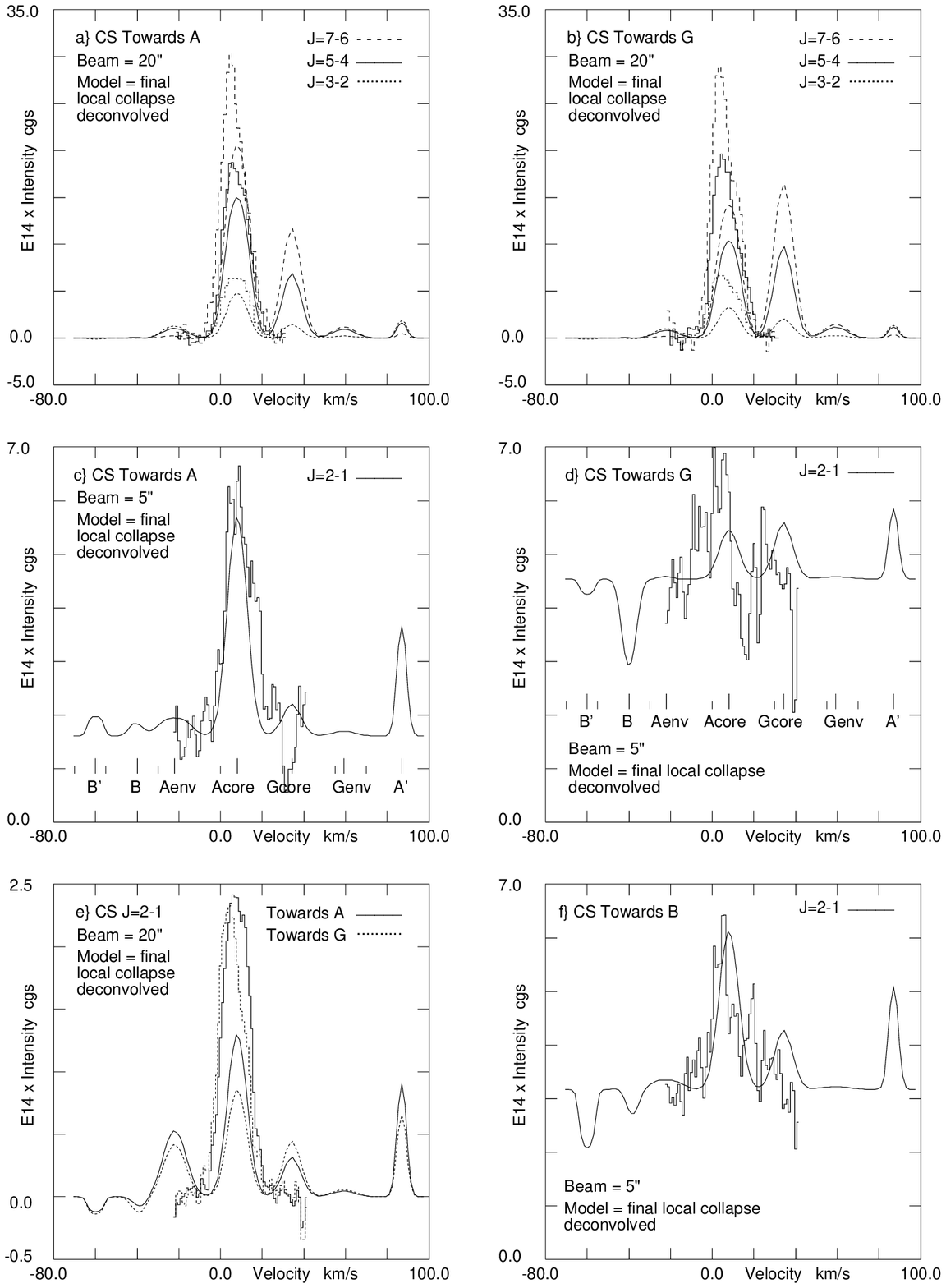}
\newpage
\plotone{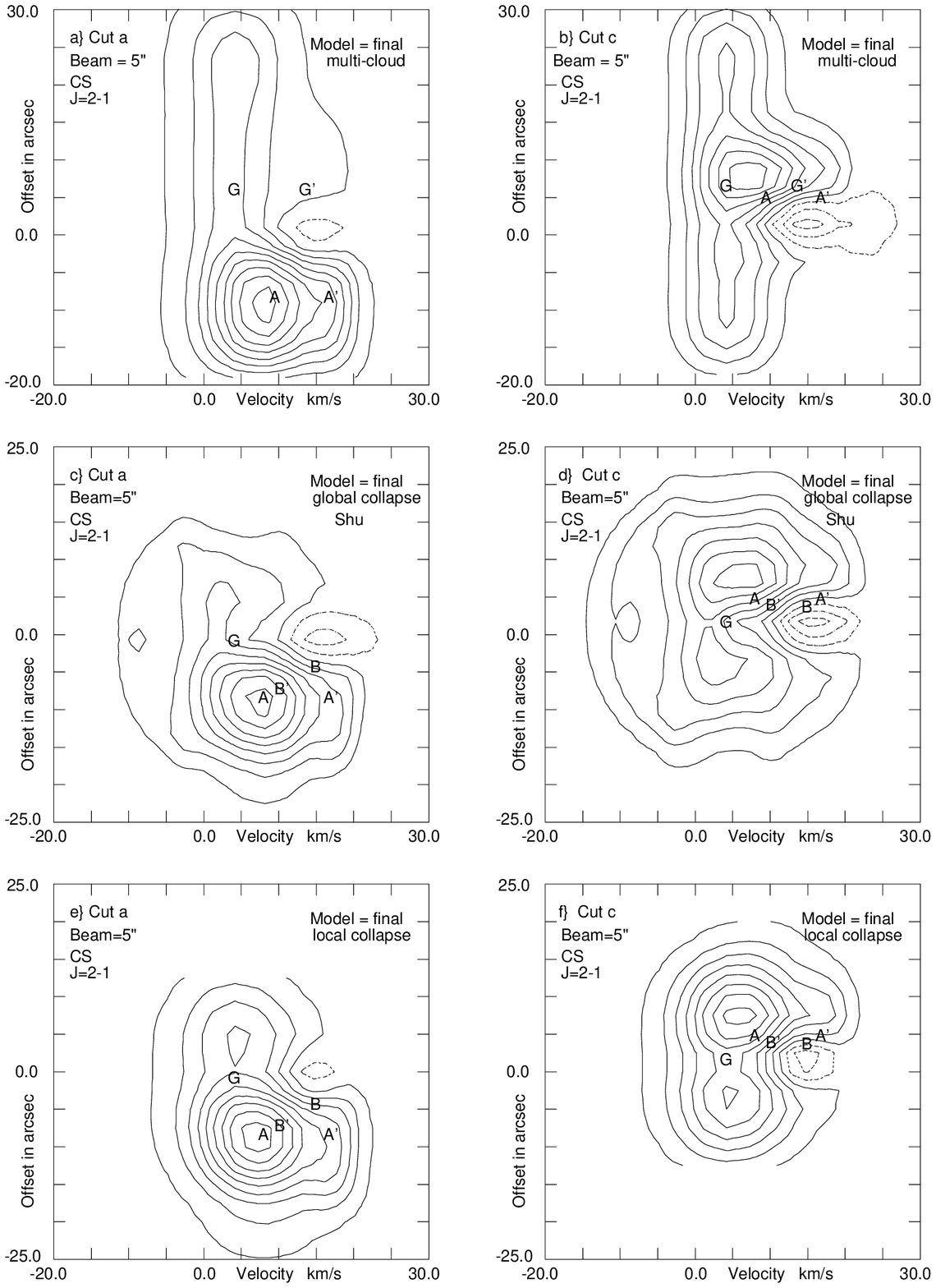}
\newpage
\plotone{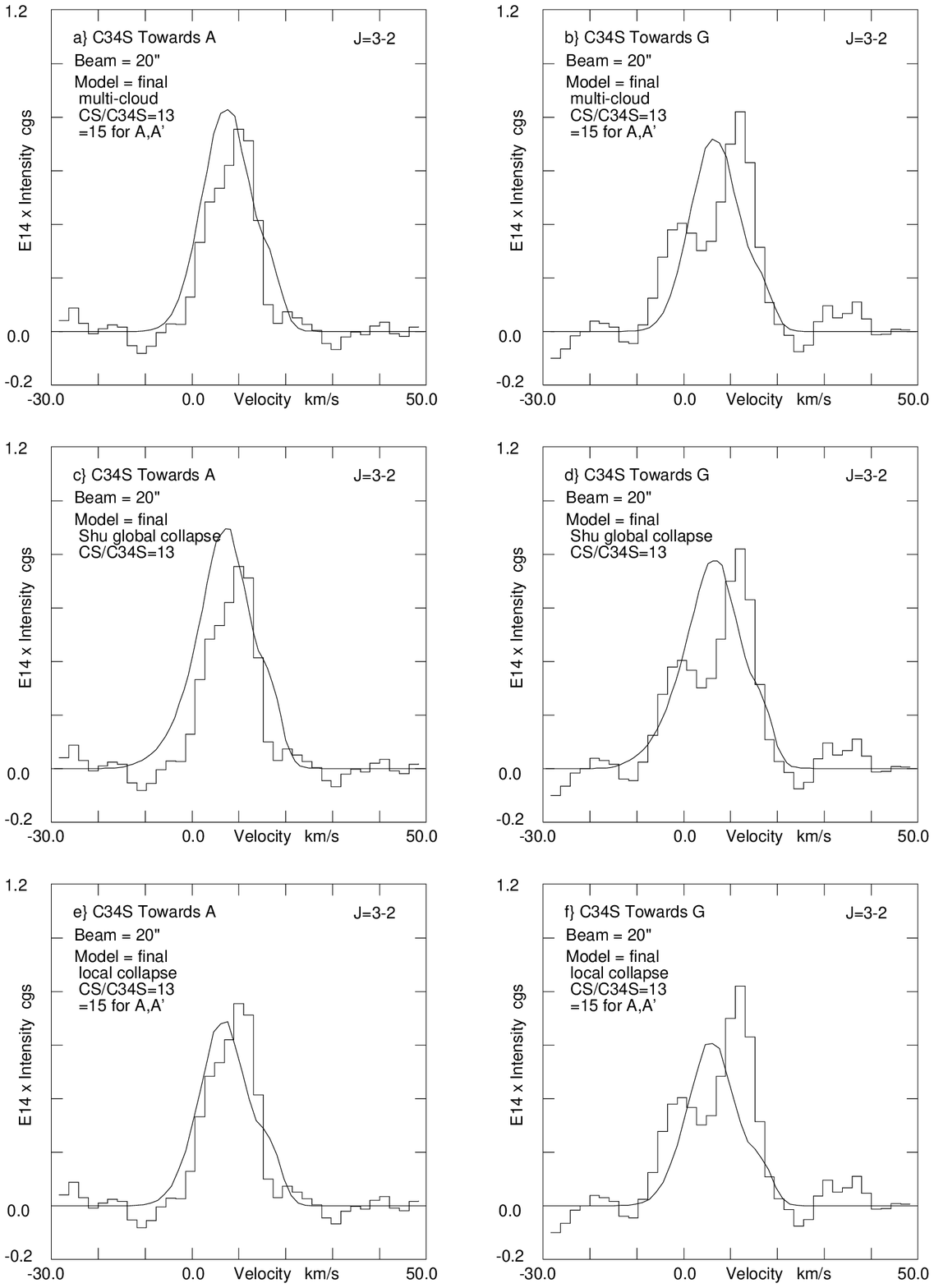}
\newpage
\plotone{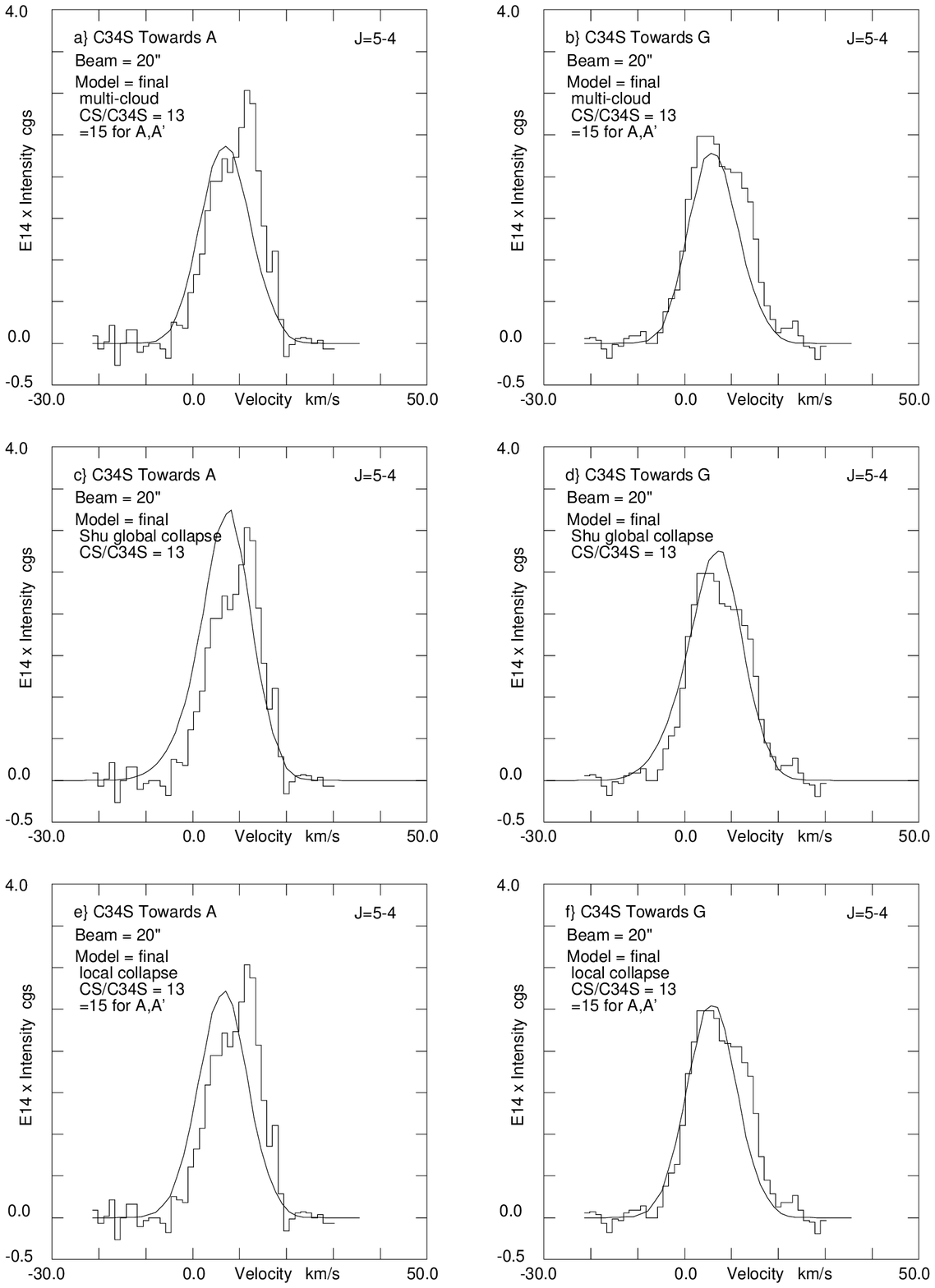}
\newpage
\plotone{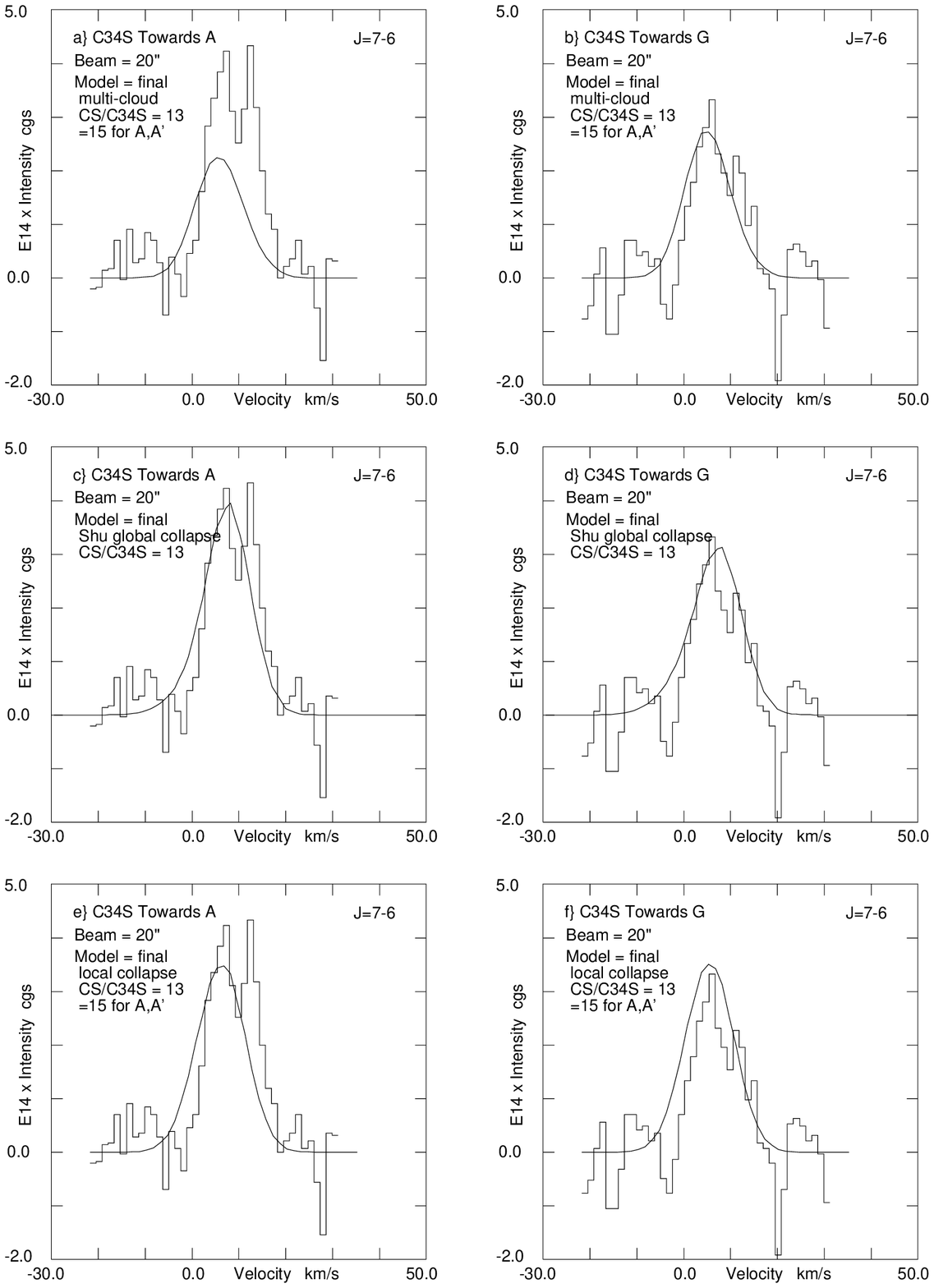}
\newpage
\plotone{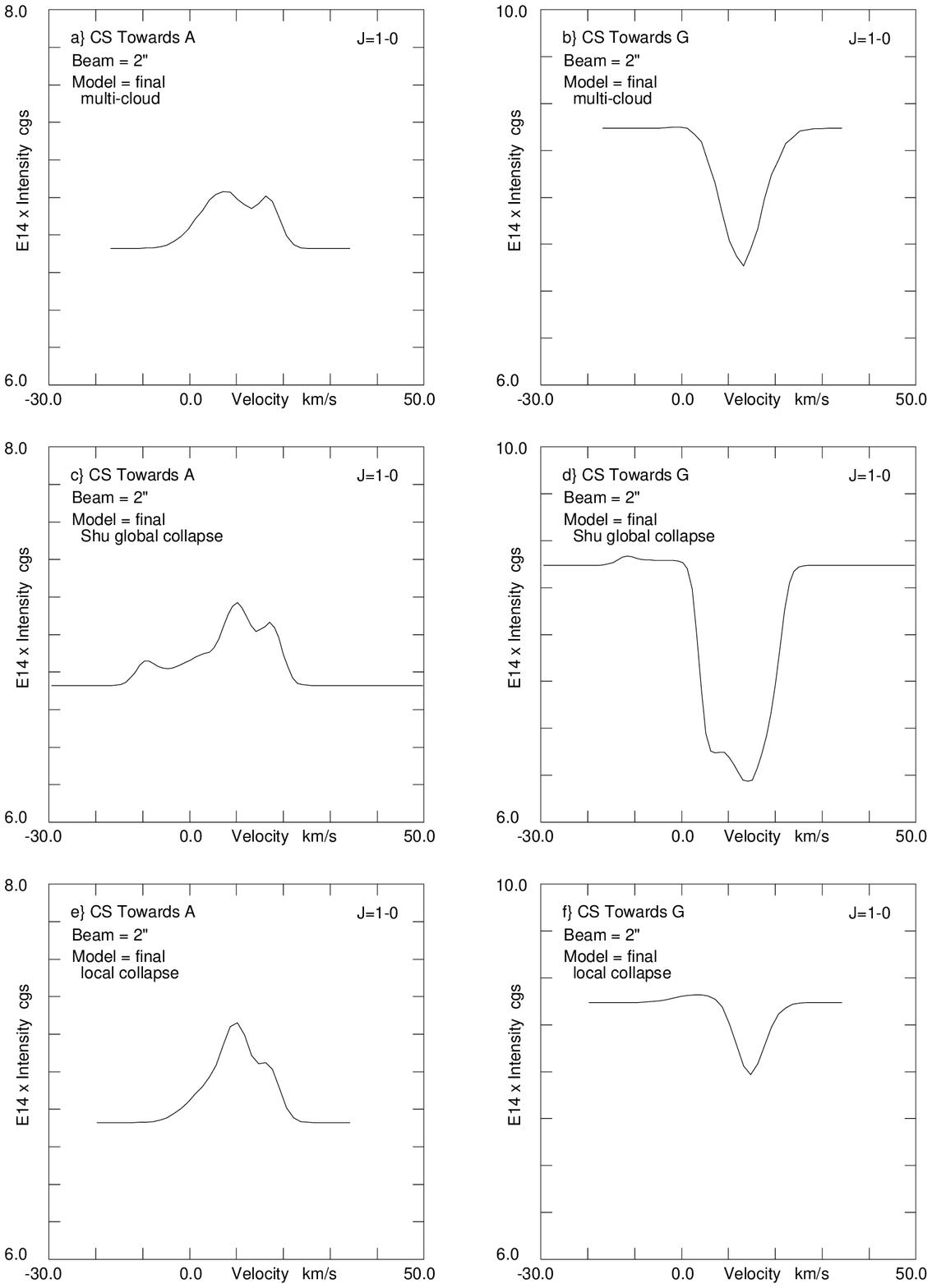}
\newpage
\plotone{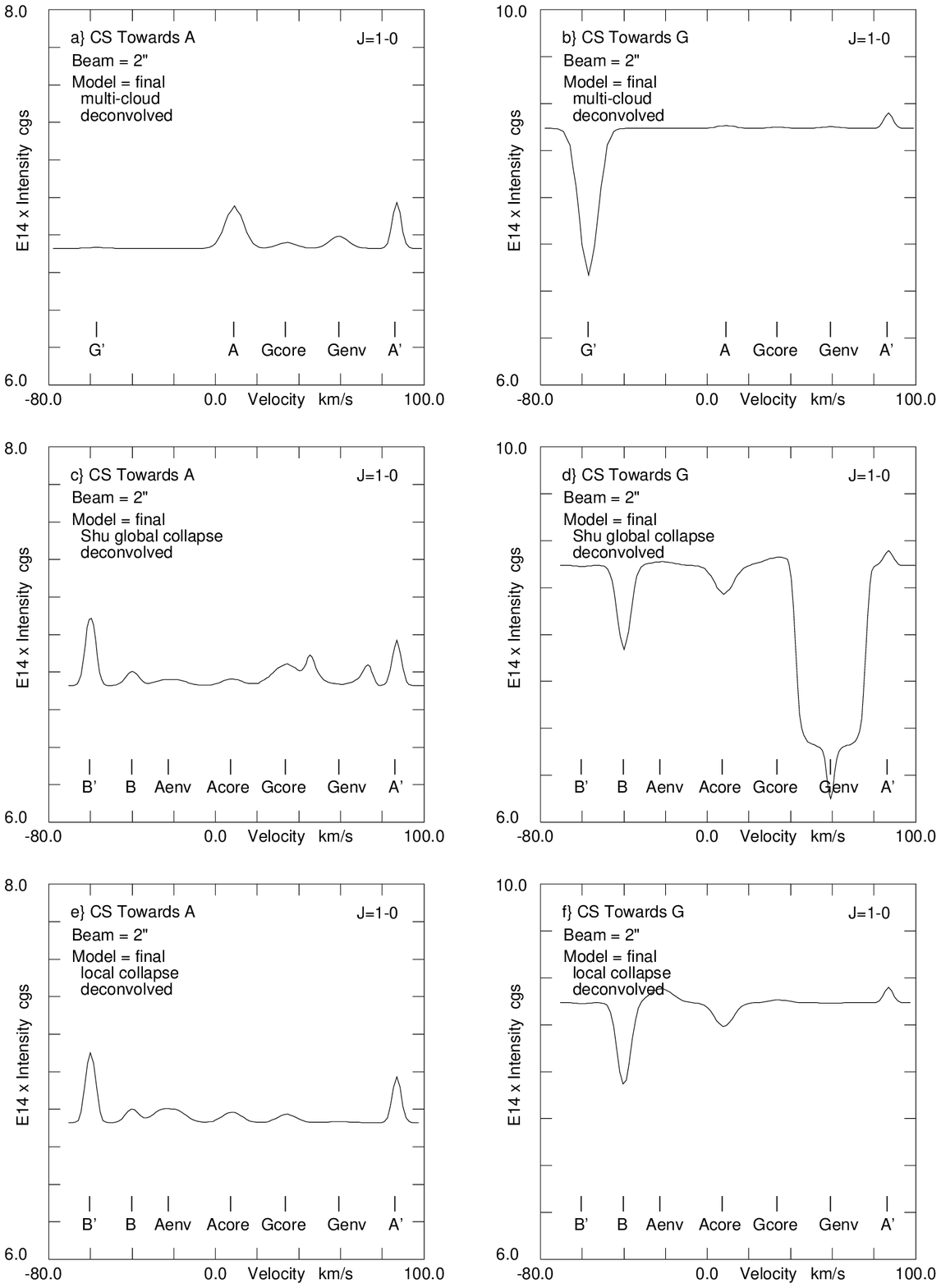}
\newpage
\plotone{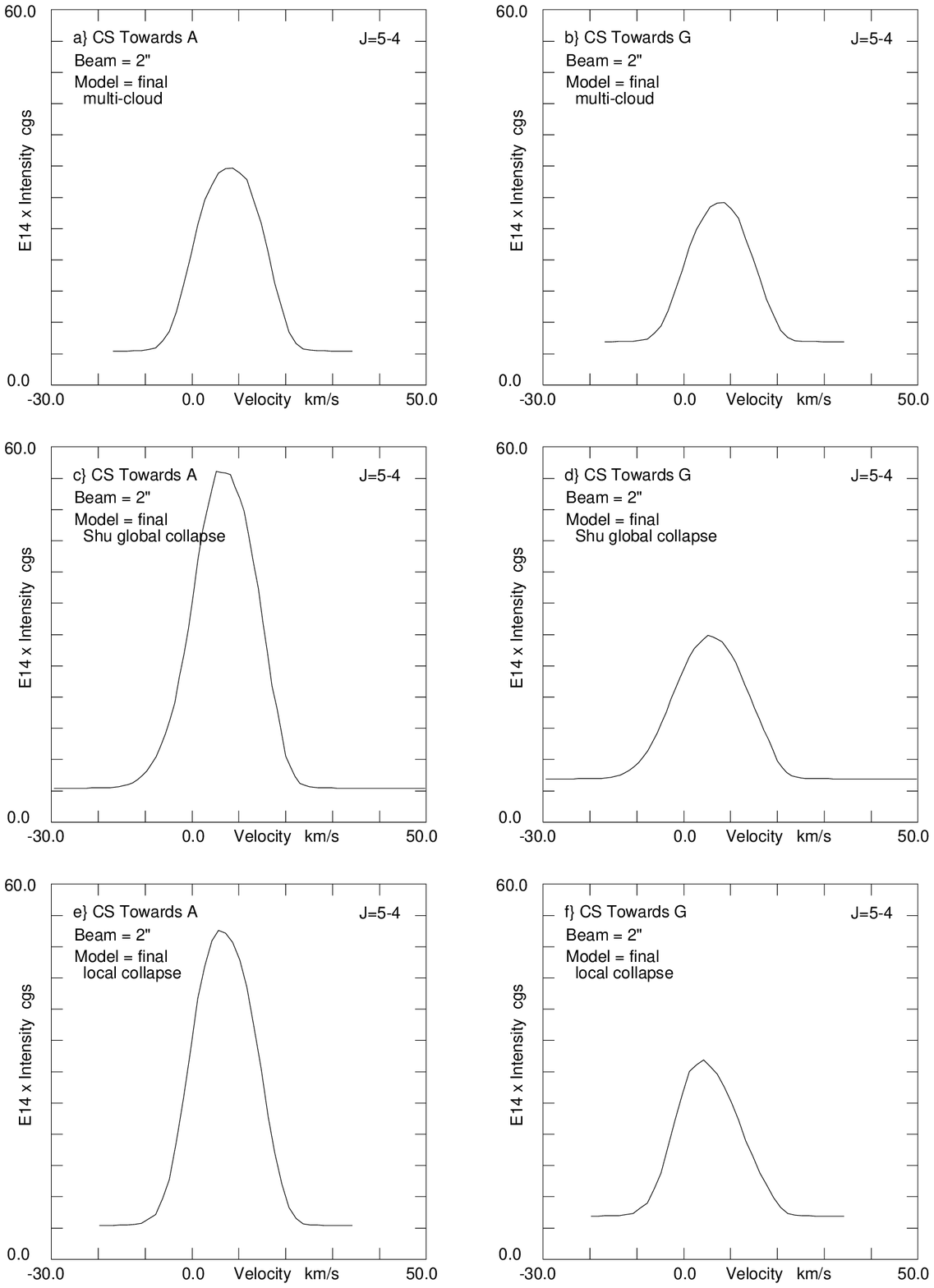}
\newpage
\plotone{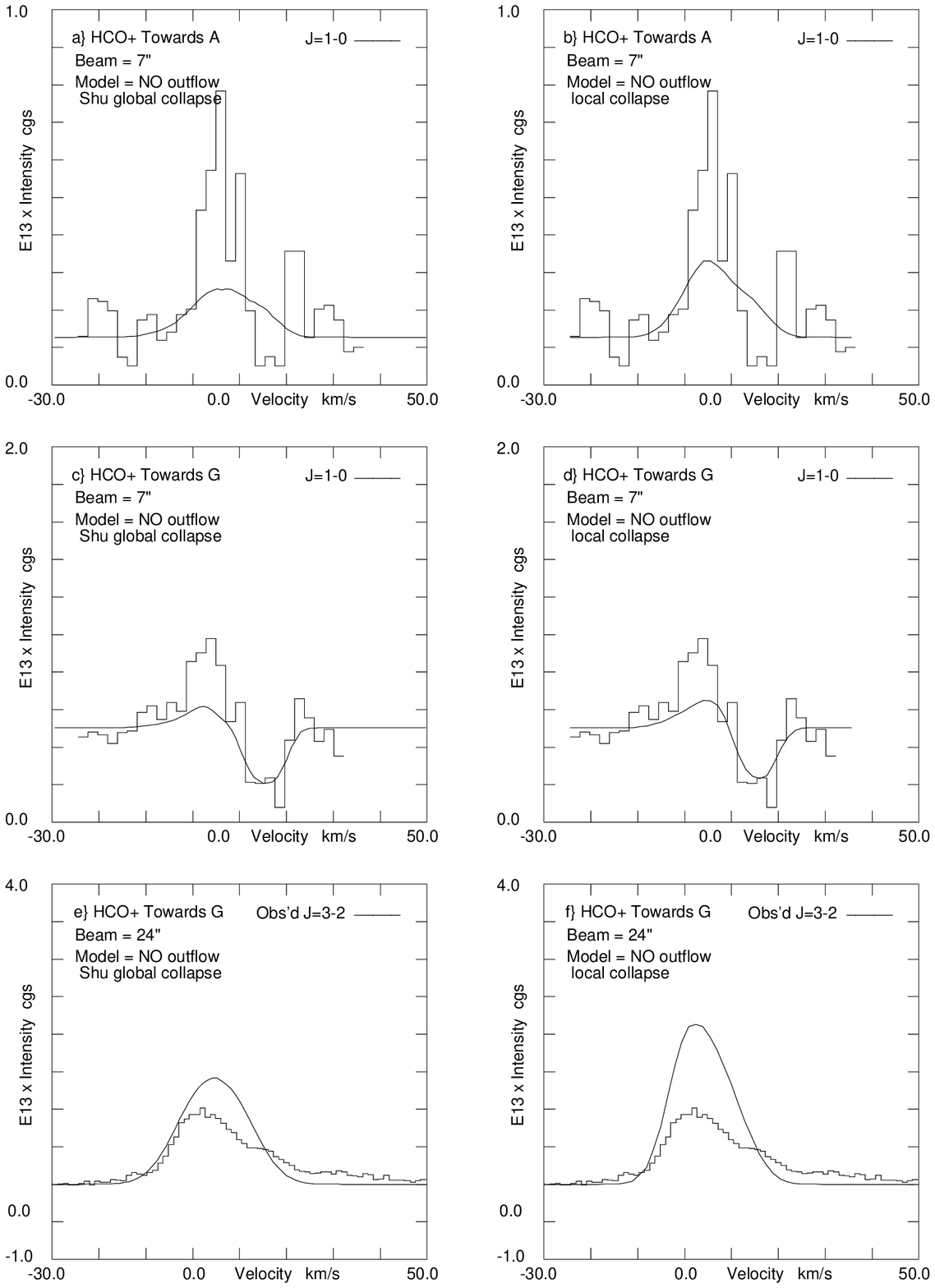}
\newpage
\plotone{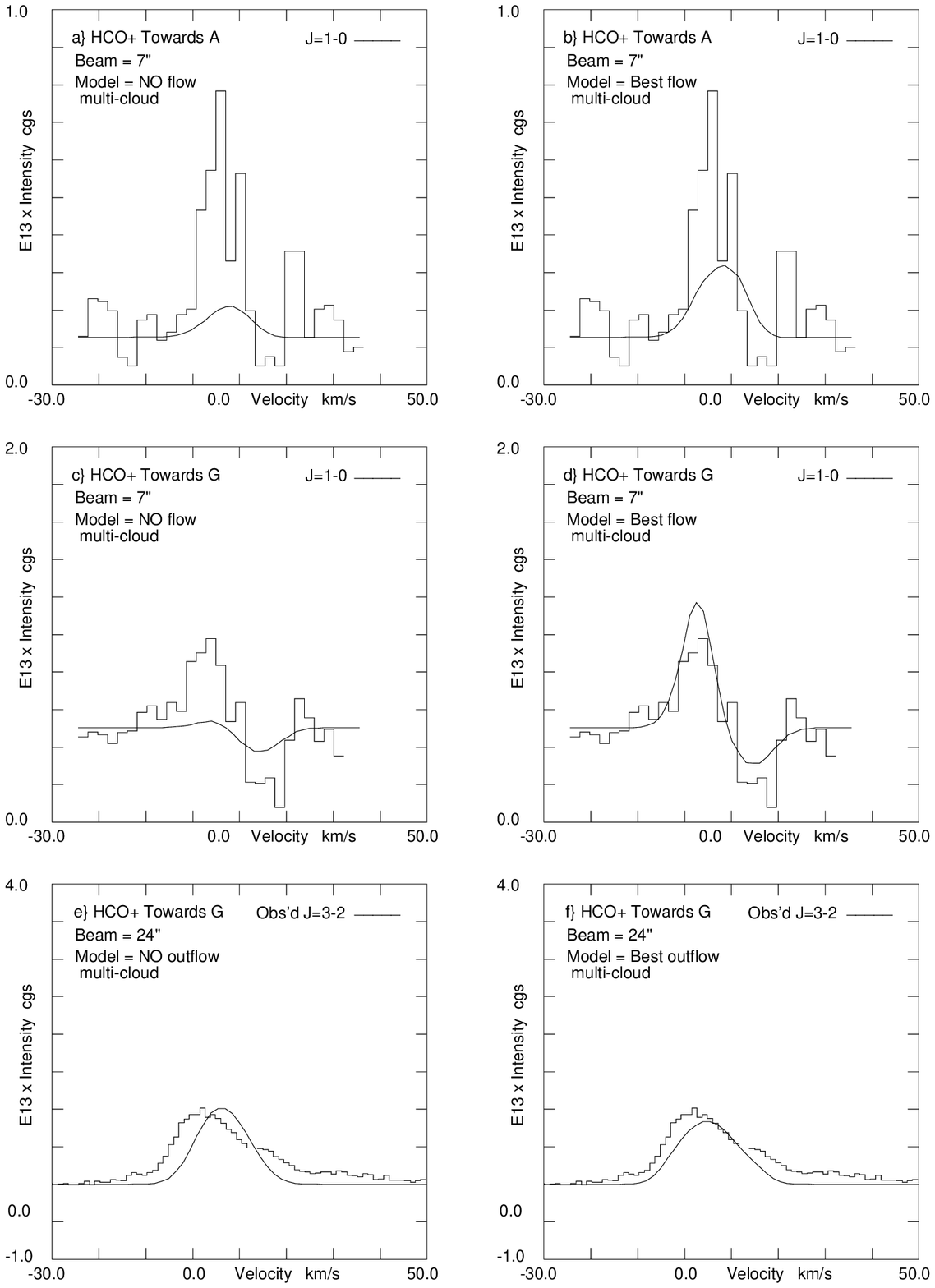}
\newpage
\plotone{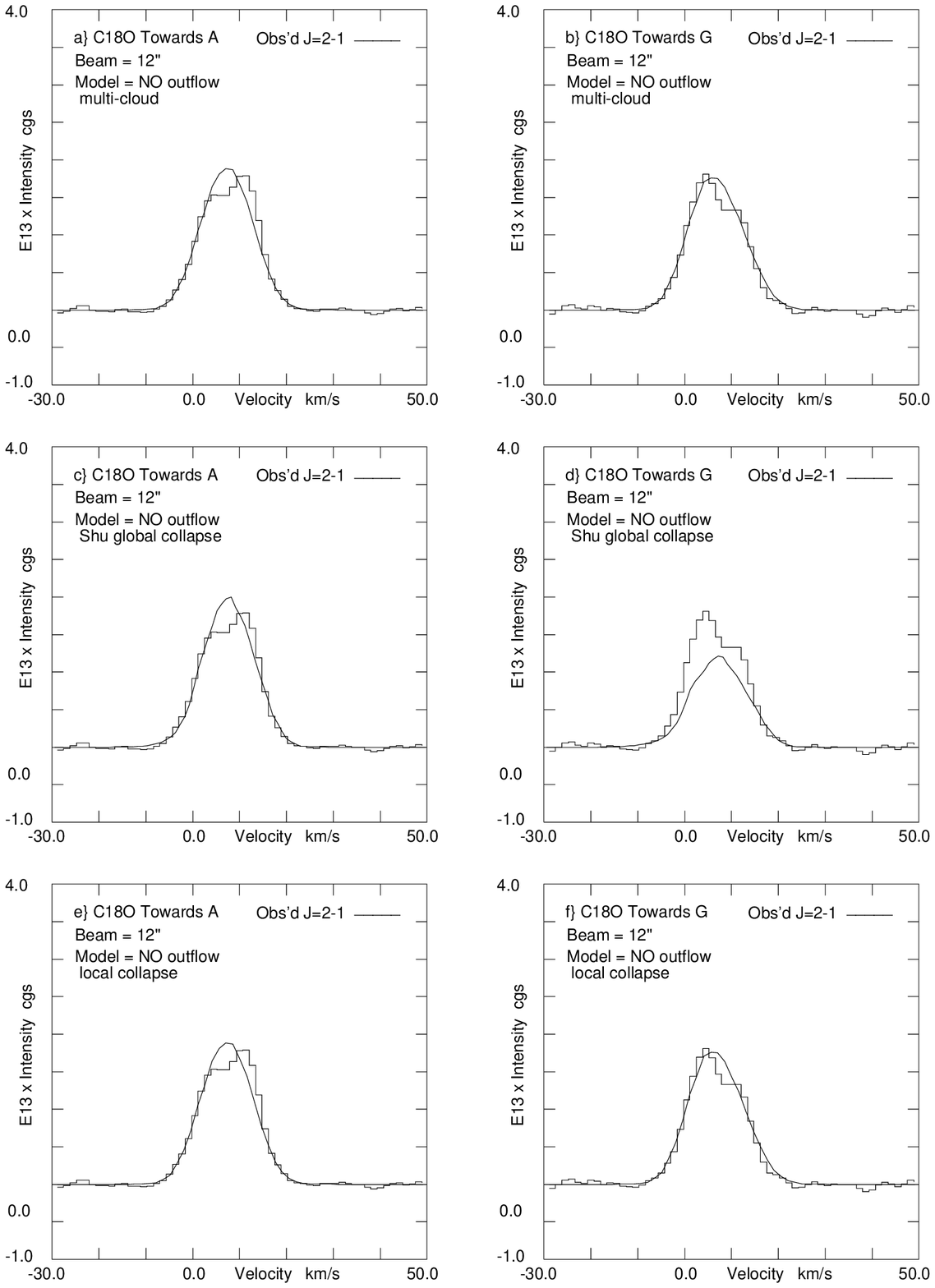}
\newpage
\plotone{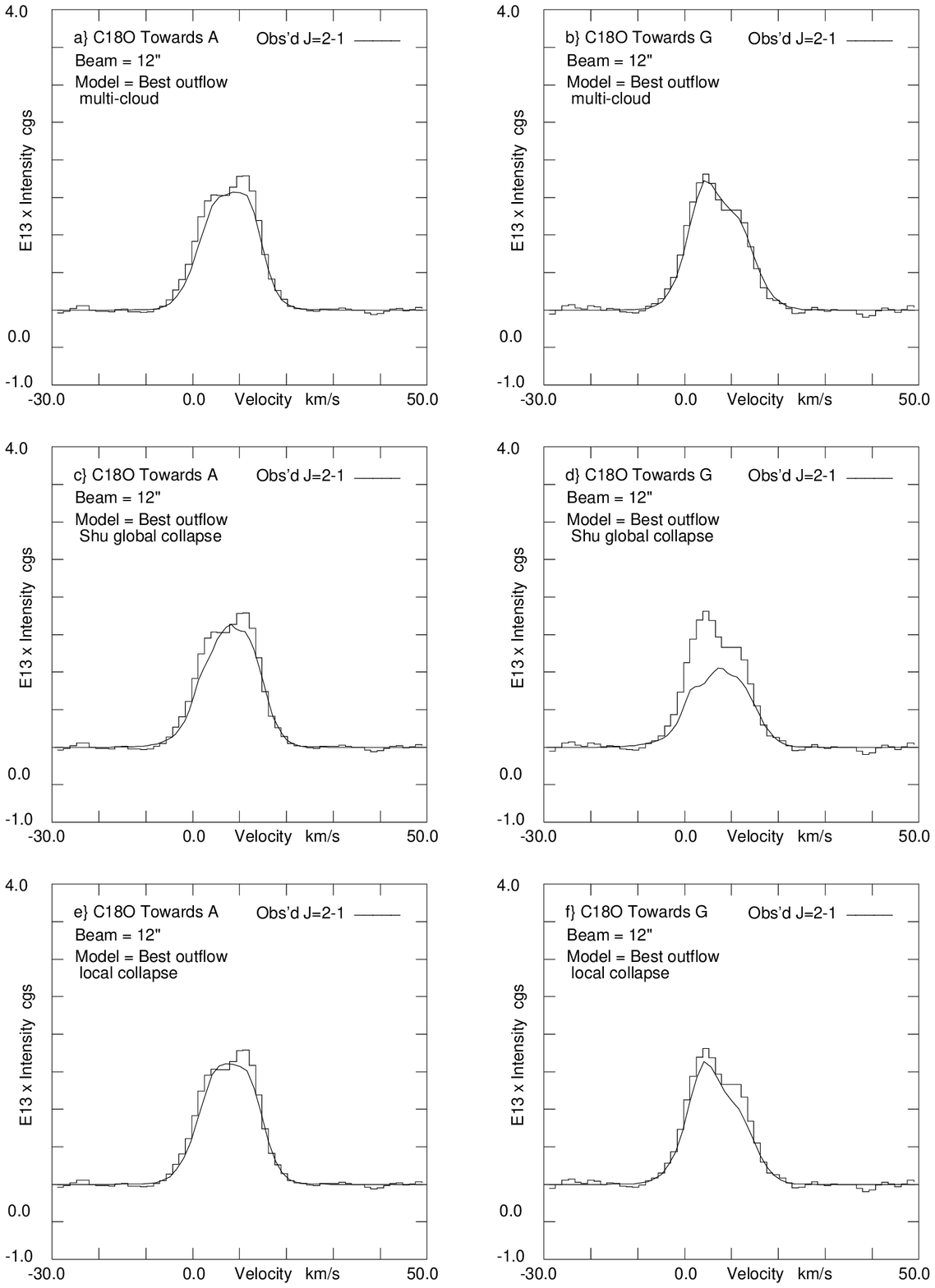}
\newpage
\plotone{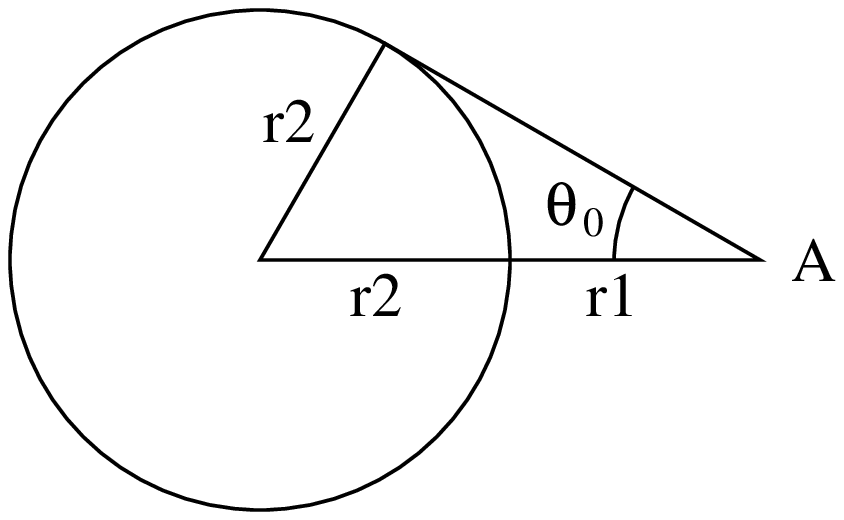}

\clearpage
%
%
\begin{deluxetable}{ccccc}
\tablecolumns{5} \tablewidth{0pc}
\tablecaption{Conversion Factor C($\nu$) between T$_B$(K) and 10$^{14}$ I(cgs)}
\tablehead{ \colhead{} & \colhead{} & \colhead{Transitions} &
\colhead{} & \colhead{} \\
\colhead{Species} & \colhead{J=2-1} & \colhead{J=3-2} &
\colhead{J=5-4} & \colhead{J=7-6} }
\startdata
CS         &   3.39   &   1.51   &   0.543   &   0.277\\
C$^{34}$S  &   3.50   &   1.56   &   0.560   &   0.286\\

\enddata
\end{deluxetable}

\clearpage
%
%
\begin{deluxetable}{lcccccc}
\tabletypesize{\small} \tablecolumns{7} \tablewidth{0pc}
\tablecaption{Preliminary Multi-cloud Model}
\tablehead{\colhead{Parameter} & \colhead{A \tablenotemark{a}} &
\colhead{G'} & \colhead{G} & \colhead{B H {\sc ii}} & \colhead{G H
{\sc ii}} & \colhead{A'}} \startdata v$_{lsr}$ (km~s$^{-1}$) &
9.0\phn & 13.0\phn\phn &  4.2\phn  & 0.0\phn \tablenotemark{b} &
0.0\phn \tablenotemark{b} &
17.0\phn\phn\\
r(H {\sc ii})~~(pc) & 0.04 & 0.01\tablenotemark{c}&
0.01\tablenotemark{c}& 0.02 & 0.17 & 0.04\tablenotemark{c}\\
n$_e$~~(cm$^{-3}$)  &    $8.6 \times 10^{4}$ &
0.10\tablenotemark{c}  & 0.10\tablenotemark{c}  &  $2.0 \times
10^{5}$ & $2.2 \times 10^{4}$ & 0.10\tablenotemark{c}\\
T$_e$~~(K) &     $10^{4}$   &   $10^{4}$    & $ 10^{4}$    &
$10^{4}$  & $10^{4}$   &   $10^{4}$\\
r$_{max}$ (pc)    &  0.60 & 0.75 & 1.43
&   0.05   & 0.33    &  0.60\\
v~~(km~s$^{-1}$) & 0.0\phn &
0.0\phn & 0.0\phn    & 0.0\phn   &   0.0\phn   & 0.0\phn\\
v$_{turb}$~~(km~s$^{-1}$) & 6.0\phn &  6.0\phn    & 6.0\phn &
4.2\phn   &   4.2\phn    & 3.0\phn\\
T$_k$~~(K) & 50.0\phn\phn & 50.0\phn\phn & 50.0\phn\phn &
50.0\phn\phn & 50.0\phn\phn &
50.0\phn\phn\\
 n(H$_2$)~~(cm$^{-3}$)   &  $1.3 \times 10^{6}$  &
$2.0 \times 10^{4}$ & $1.2 \times 10^{6}$ & $ 2.0 \times 10^{4}$ &
$2.0 \times 10^{6}$  & $2.0 \times 10^{5}$\\
n(CS)~~(cm$^{-3}$) & $ 3.4 \times 10^{-4}$  & $3.2 \times 10^{-5}$
& $6.1 \times 10^{-5}$ & $1.0 \times 10^{-7}$  & $1.0 \times
10^{-7}$ & $5.2 \times 10^{-5}$\\
mass~~$(10^{5}$ M$_\sun$)
\tablenotemark{d}&
                0.8\phn   &    0.0\phn  & 10.2\phn\phn &   0.0\phn
                &   0.2\phn   &   0.1\phn\\
\enddata
\tablenotetext{a}{H {\sc ii} region A is in the center of cloud
A.}
\tablenotetext{b}{A value of zero is used for the LSR
velocities of H {\sc ii} regions and the envelopes around them
that do not contribute to the molecular intensities.  These
velocities are not meant to imply the actual velocities of the H
{\sc ii} regions.}
\tablenotetext{c}{For clouds without H {\sc ii}
regions, very low values of r(H {\sc ii}) and ne are used in the
calculations since the rt program requires clouds to have H {\sc
ii} regions in their centers.}
\tablenotetext{d}{The total mass
for this model is $11.3 \times 10^{5}$ M$_\sun$.  The masses
include contributions from helium assuming that the number of
helium atoms is one-tenth the number of hydrogen atoms.}
\end{deluxetable}
\clearpage
%
%
\begin{deluxetable}{lcccccccccccc}
\tablecolumns{7} \tablewidth{0pc} \tablecaption{Clouds in Common
for the Final Models} \tablehead{ \colhead{} & \colhead{B'} &
\colhead{} & \multicolumn{2}{c}{B\tablenotemark{a}} &
\colhead{} & \colhead{A'}\\
\cline{2-2} \cline{4-5} \cline{7-7}\\
\colhead{Parameter\tablenotemark{b}} & \colhead{} & \colhead{} & \colhead{coef} &
\colhead{$\beta$} &
\colhead{} & \colhead{} } \startdata v$_{lsr}$~~(km~s$^{-1}$) & 10.0\phn\phn & &
15.0\phn\phn & & &
17.0\phn\phn\\
r(H {\sc ii})~~(pc) & 0.01 & & 0.02 & & & 0.04 \\
n$_e$~~(cm$^{-3}$)  & 0.10 & & $2.0 \times 10^{5}$ & & & 0.10 \\
T$_e$~~(K)    & $10^{4}$   & & $ 10^{4}$ & & & $10^4$\\
r$_{max}$~~(pc) & 0.22 & & 0.48 & & & 0.60\\
v~~(km~s$^{-1}$) & 0.0\phn & & -5.0\phn & -0.5\phn & & 0.0\phn\\
v$_{turb}$~~(km~s$^{-1}$) & 3.0\phn & & 4.0\phn & 0.0\phn & & 3.0\phn\\
T$_k$~~(K)      &
 50.0\phn\phn & & 100.0\phn\phn\phn & -0.4\phn & & 50.0\phn\phn\\
n(H$_2$)~~(cm$^{-3}$)   & $ 2.0 \times 10^{4}$ & &
$1.0 \times 10^{6}$ & -1.5 & & $2.0 \times 10^{5}$\\
n(CS)~~(cm$^{-3}$)   & $3.2 \times 10^{-4}$ & & $1.0 \times 10^{-3}$ &
-1.5\phn & & $5.2 \times 10^{-5}$\\
mass~~$(10^{5}$ M$_\sun$) & 0.0\phn & & 0.0\phn & & & 0.1\phn\\
\enddata
\tablenotetext{a}{H~{\sc ii} region B is at the center of cloud B.}
\tablenotetext{b}{Some of the parameters have gradients in the form of power laws,
where $\beta$ is the exponent in coefficient$\times$(r/r(H {\sc ii})$^{\beta}$.  If
$\beta$ for the core is zero in core-envelope models, then the expression is
coefficient$\times$(r/r$_{max})^{\beta}$, where r$_{max}$ is the radius of the
core.}
\end{deluxetable}
\clearpage
%
%
\begin{deluxetable}{lccccccc}
\tabletypesize{\small} \tablecolumns{7} \tablewidth{0pc}
\tablecaption{Final Multi-cloud Model\tablenotemark{a}}
\tablehead{\colhead{Parameter} & \colhead{A \tablenotemark{b}} &
\colhead{G'} & \colhead{G} & \colhead{G env} & \colhead{B H {\sc
ii}} & \colhead{G H {\sc ii}}} \startdata v$_{lsr}$ (km~s$^{-1}$)
& 9.0\phn & 13.0\phn\phn & 4.2\phn & & 0.0\phn \tablenotemark{c} &
0.0\phn
\tablenotemark{c}\\
r(H {\sc ii}) (pc) & 0.04 &
0.01\tablenotemark{d}& 0.01\tablenotemark{d}&
 & 0.02 & 0.17 \\
n$_e$ (cm$^{-3}$)  &    $8.6 \times 10^{4}$ &
0.10\tablenotemark{d}  & 0.10\tablenotemark{d} & &  $2.0 \times
10^{5}$ & $2.2 \times 10^{4}$ \\
T$_e$ (K) & $10^{4}$ & $10^{4}$ & $ 10^{4}$ &  & $10^{4}$ &
$10^{4}$ \\
r$_{max}$ (pc) &  0.60 & 0.75 & 1.01 & 1.43 & 0.05 & 0.33 \\
v~~(km~s$^{-1}$) & 0.0\phn & 0.0\phn & 0.0\phn & 0.0\phn & 0.0\phn
& 0.0\phn \\ v$_{turb}$~~(km~s$^{-1}$) & 6.0\phn & 6.0\phn &
6.0\phn & 6.0\phn & 4.2\phn & 4.2\phn\\ T$_k$~~(K) & 50.0\phn\phn
& 50.0\phn\phn & 100.0\phn\phn\phn
 &  50.0\phn\phn &  50.0\phn\phn &  50.0\phn\phn \\
n(H$_2$)~~(cm$^{-3}$)   &  $1.3 \times 10^{6}$  & $2.0 \times 10^{4}$ & $6.0 \times
10^{6}$ & $1.2 \times 10^{6}$ & $ 2.0 \times 10^{4}$ & $2.0 \times 10^{6}$ \\
n(CS)~~(cm$^{-3}$)   &  $ 3.4 \times 10^{-4}$  & $3.2 \times 10^{-5}$ & $6.1 \times
10^{-5}$ & $6.1 \times 10^{-5}$ & $1.0 \times 10^{-7}$  & $1.0 \times 10^{-7}$ \\
mass~~$(10^{5}$ M$_\sun$) \tablenotemark{e}& 0.8\phn & 0.0\phn & 17.9\phn\phn &
6.6\phn & 0.0\phn & 0.2\phn\\
\enddata
\tablenotetext{a}{In addition to the clouds in this table, the
final multi-cloud model includes cloud A' from Table 3 after cloud
G H {\sc ii}.} \tablenotetext{b}{H {\sc ii} region A is in the
center of cloud A.} \tablenotetext{c}{A value of zero is used for
the LSR velocities of H {\sc ii} regions and the envelopes around
them that do not contribute to the molecular intensities.  These
velocities are not meant to imply the actual velocities of the H
{\sc ii} regions.} \tablenotetext{d}{For clouds without H {\sc ii}
regions, very low values of r(H {\sc ii}) and ne are used in the
calculations since the rt program requires clouds to have H {\sc
ii} regions in their centers.} \tablenotetext{e}{The total mass
for this model is $25.6 \times 10^{5}$ M$_\sun$.  The masses
include contributions from helium assuming that the number of
helium atoms is one-tenth the number of hydrogen atoms.}
\end{deluxetable}
\clearpage
%
%
\begin{deluxetable}{lcccccccccccc}
\tabletypesize{\footnotesize} \tablecolumns{12} \tablewidth{7in}
\tablecaption{Final Global Collapse Model \tablenotemark{a}}
\tablehead{ \colhead{} & \multicolumn{2}{c}{A
core\tablenotemark{b}}& \colhead{} & \multicolumn{2}{c}{A env} &
\colhead{} & \multicolumn{2}{c}{G core\tablenotemark{b}} &
\colhead{} &
\multicolumn{2}{c}{G env} \\
\cline{2-3} \cline{5-6} \cline{8-9} \cline{11-12}\\
 \colhead{Parameter\tablenotemark{c}} & \colhead{coef} & \colhead{$\beta$} & \colhead{} &
\colhead{coef} & \colhead{$\beta$} & \colhead{} & \colhead{coef} & \colhead{$\beta$}
& \colhead{} & \colhead{coef} & \colhead{$\beta$} & }
\startdata
v$_{lsr}$~~(km~s$^{-1}$) & 8.0\phn & & & & & &
4.2\phn & & & & \\
r(H {\sc ii})~~(pc) & 0.04 & & & & & &
0.17 & & & & \\
n$_e$~~(cm$^{-3}$)  & $8.6 \times 10^{4}$ & & & & & &
$2.2 \times 10^{4}$ & & & & \\
T$_e$~~(K)    & $10^{4}$ & & & & & &
$10^{4}$& & & & \\
r$_{max}$~~(pc) & 0.48 & & & 0.96 & & &
0.81 & & & 4.97 & \\
v~~(km~s$^{-1}$) & 0.0\phn & 0.0\phn & & -5.0\phn & -0.5\phn & &
0.0 & 0.0 & & -22.5\phn\phn & -0.5\phn \\
v$_{turb}$~~(km~s$^{-1}$) & 6.0\phn & 0.0\phn & & 6.0\phn & 0.0\phn & &
9.0\phn & 0.0\phn & & 2.0 & 0.0\phn \\
T$_k$~~(K)      & 100.0\phn\phn\phn & -0.4\phn & & 100.0\phn\phn\phn &
-0.4\phn & & 100.0\phn\phn\phn & -0.4\phn & & 100.0\phn\phn\phn & -0.4\phn \\
n(H$_2$)~~(cm$^{-3}$)   &  $8.0 \times 10^{6}$ & 0.0\phn & &
$8.0 \times 10^{5}$  & -1.5\phn & &
$1.5 \times 10^{6}$ & 0.0\phn & & $1.5 \times 10^{3}$ & -1.5 \\
n(CS)~~(cm$^{-3}$)   &  $ 7.0 \times 10^{-4}$ & 0.0\phn & & $7.0 \times 10^{-5}$ &
-1.5 & &
$1.6 \times 10^{-4}$ & 0.0\phn & & $1.6 \times 10^{-5}$  & -1.5\phn \\
mass~~$(10^{5}$ M$_\sun$) \tablenotemark{d}&
2.6\phn   & & &  0.9\phn  & & & 2.3\phn & & & 0.0\phn & \\
\enddata
\tablenotetext{a}{In addition to the clouds in this table, the
final global collapse model includes clouds B', B, and A' from
Table 3.  Clouds B' and B are between clouds A and G, and cloud A'
is behind cloud G.} \tablenotetext{b}{H~{\sc ii} region A is in
the center of cloud A, and H~{\sc ii} region G is in the center of
cloud G.} \tablenotetext{c}{ Some of the parameters have gradients
in the form of power laws, where $\beta$ is the exponent in
coefficient$\times$(r/r(H {\sc ii})$^{\beta}$.  If $\beta$ for the
core is zero in core-envelope models, then the expression is
coefficient$\times$(r/r$_{max})^{\beta}$, where r$_{max}$ is the
radius of the core.} \tablenotetext{d}{The total mass for this
model is $5.9 \times 10^{5}$ M$_\sun$.  The masses include
contributions from helium assuming that the number of helium atoms
is one-tenth the number of hydrogen atoms.}
\end{deluxetable}
\clearpage
%
%
\begin{deluxetable}{lcccccccccccc}
\tabletypesize{\footnotesize} \tablecolumns{12} \tablewidth{7in}
\tablecaption{Final Local Collapse Model \tablenotemark{a}}
\tablehead{ \colhead{} & \multicolumn{2}{c}{A
core\tablenotemark{b}}& \colhead{} & \multicolumn{2}{c}{A env} &
\colhead{} & \multicolumn{2}{c}{G core\tablenotemark{b}} &
\colhead{} &
\multicolumn{2}{c}{G env} \\
\cline{2-3} \cline{5-6} \cline{8-9} \cline{11-12}\\
 \colhead{Parameter\tablenotemark{c}} & \colhead{coef} & \colhead{$\beta$} & \colhead{} &
\colhead{coef} & \colhead{$\beta$} & \colhead{} & \colhead{coef} & \colhead{$\beta$}
& \colhead{} & \colhead{coef} & \colhead{$\beta$} & } \startdata
v$_{lsr}$~~(km~s$^{-1}$) & 8.0\phn & & & & & &
4.2\phn & & & & \\
r(H {\sc ii})~~(pc) & 0.04 & & & & & &
0.17 & & & & \\
n$_e$~~(cm$^{-3}$)  & $8.6 \times 10^{4}$ & & & & & &
$2.2 \times 10^{4}$ & & & & \\
T$_e$~~(K)    & $10^{4}$ & & & & & &
$10^{4}$& & & & \\
r$_{max}$~~(pc) & 0.48 & & & 0.96 & & &
0.52 & & & 0.79 & \\
v~~(km~s$^{-1}$) & 0.0\phn & 0.0\phn & & -5.0\phn & -0.5\phn & &
0.0 & 0.0 & & -5.0\phn & -0.5\phn \\
v$_{turb}$~~(km~s$^{-1}$) & 6.0\phn & 0.0\phn & & 6.0\phn & 0.0\phn & &
6.0\phn & 0.0\phn & & 6.0\phn & 0.0\phn \\
T$_k$~~(K)      & 100.0\phn\phn\phn & -0.4\phn & & 100.0\phn\phn\phn &
-0.4\phn & & 100.0\phn\phn\phn & -0.4\phn & & 100.0\phn\phn\phn & -0.4\phn \\
n(H$_2$)~~(cm$^{-3}$)   &  $5.0 \times 10^{6}$ & 0.0\phn & &
$5.0 \times 10^{5}$  & -1.5\phn & &
$5.0 \times 10^{6}$ & 0.0\phn & & $5.0 \times 10^{6}$ & -1.5 \\
n(CS)~~(cm$^{-3}$)   &  $ 6.0 \times 10^{-4}$ & 0.0\phn & & $6.0 \times 10^{-5}$ &
-1.5 & &
$1.8 \times 10^{-4}$ & 0.0\phn & & $1.8 \times 10^{-4}$  & -1.5\phn \\
mass~~$(10^{5}$ M$_\sun$) \tablenotemark{d}&
1.6\phn   & & &  0.6\phn  & & & 2.0\phn & & & 3.5\phn & \\
\enddata
\tablenotetext{a}{In addition to the clouds in this table, the
final local collapse model includes clouds B', B, and A' from
Table 3.  Clouds B' and B are between clouds A and G, and cloud A'
is behind cloud G.} \tablenotetext{b}{H~{\sc ii} region A is in
the center of cloud A, and H~{\sc ii} region G is in the center of
cloud G.} \tablenotetext{c}{ Some of the parameters have gradients
in the form of power laws, where $\beta$ is the exponent in
coefficient$\times$(r/r(H {\sc ii})$^{\beta}$.  If $\beta$ for the
core is zero in core-envelope models, then the expression is
coefficient$\times$(r/r$_{max})^{\beta}$, where r$_{max}$ is the
radius of the core.} \tablenotetext{d}{The total mass for this
model is $7.8 \times 10^{5}$ M$_\sun$.  The masses include
contributions from helium assuming that the number of helium atoms
is one-tenth the number of hydrogen atoms.}
\end{deluxetable}
%
%
\clearpage
\begin{deluxetable}{cccccccc}
\tablecolumns{8} \tablewidth{0pc} \tabletypesize{\footnotesize}
\tablecaption{The RMS Differences Between Model Profiles and
Observed Profiles} \tablehead{ \colhead{} & \colhead{} &
\colhead{} & \colhead{} & \colhead{Pre-M-C\tablenotemark{b}} &
\colhead{Multi-Cloud} & \colhead{Global Collapse} & \colhead{Local Collapse}\\
\colhead{Region} & \colhead{Resolution} & \colhead{Transition} &
\colhead{$\sigma_{base}$\tablenotemark{a}} &
\colhead{$\sigma_{line}/\sigma_{base}$} & \colhead{$\sigma_{line}/\sigma_{base}$} &
\colhead{$\sigma_{line}/\sigma_{base}$} & \colhead{$\sigma_{line}/\sigma_{base}$} }
\startdata
A      &    5     &   2-1  &   4.1 & 1.0 & 1.2 &  1.6 &  1.4\\
G      &    5     &   2-1  &   4.1 & 1.5 & 1.7 &  1.7 &  1.5\\
B      &    5     &   2-1  &   4.1 & 1.9 & 2.0 &  1.7 &  1.5\\
A      &   20     &   2-1  &   0.8 & 3.4 & 4.1 &  4.4 &  6.1\\
G      &   20     &   2-1  &   0.8 & 3.2 & 5.3 &  4.3 &  6.5\\
A      &   20     &   3-2  &   1.2 & 9.3 & 7.5 &  6.9 &  5.3\\
G      &   20     &   3-2  &   1.2 & 5.6 & 4.3 &  4.8 &  7.3\\
A      &   20     &   5-4  &   6.5 & 2.0 & 3.2 &  4.5 &  3.7\\
G      &   20     &   5-4  &   6.5 & 1.9 & 3.4 &  4.1 &  3.1\\
A      &   20     &   7-6  &   11\phn\phn & 7.1 & 2.4 &  3.3 & 2.5\\
G      &   20     &   7-6  &   11\phn\phn & 7.6 & 2.7 &  3.4 & 3.7\\
\tableline
average &         &        &       & 4.0 & 3.4 &  3.7 &  3.9\\
\enddata
\tablecomments{All $\sigma$ values are times 10$^{-15}$ erg
s$^{-1}$ cm$^{-2}$ Hz$^{-1}$ sr$^{-1}$} \tablenotetext{a}{rms
variations of the observations of the line-free baseline or
continuum.} \tablenotetext{b}{Preliminary Multi-Cloud Model}
\end{deluxetable}
%
%
\clearpage
\begin{deluxetable}{cccc}
\tablecolumns{4} \tablewidth{0pc}
\tablecaption{Percent of the Celestial Sphere Covered}
\tablehead{ \colhead{ r\tablenotemark{a}} & \colhead{$\theta$} &
\colhead{$\theta$} &
\colhead{Percent of Celestial}\\
\colhead{} & \colhead{(radians)} & \colhead{(degrees)} & \colhead{Sphere Covered by}\\
\colhead{} & \colhead{}          & \colhead{}          & \colhead{External Cloud}
}
\startdata
                              0   & 0        &   0             &    0\\
                              0.1 & 0.091    &   5.2           &    0.2\\
                              0.5 & 0.340    &  19.5\phn       &    2.9\\
                              1.0 & 0.524    &  30.0\phn       &    6.7\\
                            2.0   & 0.730    &  41.8\phn       &    12.7\phn\\
                        10.0\phn  & 1.141    &  65.4\phn       &    29.2\phn\\
                         $\infty$ & 1.571    &  90.0\phn       &    50.0\phn\\
\enddata
\tablenotetext{a}{$r=r_2/r_1$ where $r_1$ is the distance from
a point within one cloud to the outer edge of a second, external
cloud whose radius is $r_2$.  (See Figure 27).}
\end{deluxetable}
\end{document}